\documentclass[final,5p,times,twocolumn]{elsarticle}

\usepackage{amssymb}
\usepackage{amsmath}

\usepackage{url}
\usepackage{multirow}
\usepackage{booktabs}
\usepackage{color}
\usepackage{amsmath}
\usepackage{colortbl}
\usepackage{graphicx}
\usepackage{amssymb}
\usepackage{comment}
\usepackage{xparse}
\usepackage{xspace}
\usepackage[caption=false,font=footnotesize,labelfont=rm,textfont=rm,subrefformat=parens,labelformat=parens]{subfig}
\usepackage{pifont}
\usepackage{fontawesome5}

\usepackage{algorithm}
\usepackage{algpseudocode}

\usepackage[table]{xcolor}

\definecolor{PrimaryRowFg}{rgb}{0.0, 0.0, 0.0}
\definecolor{SecondaryRowFg}{rgb}{0.0, 0.0, 0.0}
\definecolor{BaselineRowBg}{gray}{0.9}
\definecolor{MissingRowBg}{gray}{0.9}
\definecolor{SelectedRowBg}{HTML}{e1eafa}

\definecolor{Blue}{HTML}{4285f4}
\definecolor{Red}{HTML}{eb3838}
\definecolor{Green}{HTML}{21bf52}

\definecolor{DarkBlue}{HTML}{155cd1}
\definecolor{DarkRed}{HTML}{c72020}
\definecolor{DarkGreen}{HTML}{119139}

\usepackage{siunitx}
\DeclareSIUnit\flop{FLOPs}

\usepackage{utfsym}
\usepackage{bm}

\usepackage{orcidlink}

\DeclareMathOperator{\cossim}{sim}
\DeclareMathOperator{\softmax}{softmax}

\newcommand{\eer}{EER (\%)}
\newcommand{\mindcf}{$\text{minDCF}_\text{0.01}$}

\NewDocumentCommand{\todo}{o}{\textcolor{Red}{(TO-DO\IfValueT{#1}{:~#1})}\xspace}

\newcommand{\cmark}{\textcolor{Green}{\ding{51}}}
\newcommand{\xmark}{\textcolor{Red}{\ding{55}}}

\newcommand{\titlecaption}[2]{\caption{\textbf{#1}. #2}}

\usepackage{etoolbox}

\newcommand{\ac}[1]{#1}
\newcommand{\acs}[1]{#1}
\newcommand{\acf}[1]{#1}

\definecolor{ReviewColor}{HTML}{144eaf}

\newcommand{\m}{\textcolor[HTML]{e6e6e6}{\raisebox{0.5ex}{\scalebox{0.35}{\faCircle}}}}

\journal{Speech Communication}

\begin{document}

\begin{frontmatter}



\title{Self-Supervised Learning for Speaker Recognition: A study and review}


\author{Theo Lepage\corref{cor1}~\orcidlink{0009-0009-0676-4099}} 
\ead{theo.lepage@epita.fr}
\cortext[cor1]{Corresponding author}

\author{Reda Dehak~\orcidlink{0000-0002-4078-7261}} 
\ead{reda.dehak@epita.fr}

\def\theaffn{} 
\affiliation[]{organization={EPITA Research Laboratory (LRE)},
            country={France}}

\begin{abstract}
Deep learning models trained in a supervised setting have revolutionized audio and speech processing. However, their performance inherently depends on the quantity of human-annotated data, making them costly to scale and prone to poor generalization under unseen conditions. To address these challenges, Self-Supervised Learning (SSL) has emerged as a promising paradigm, leveraging vast amounts of unlabeled data to learn relevant representations. The application of SSL for Automatic Speech Recognition (ASR) has been extensively studied, but research on other downstream tasks, notably Speaker Recognition (SR), remains in its early stages. This work describes major SSL instance-invariance frameworks (e.g., SimCLR, MoCo, and DINO), initially developed for computer vision, along with their adaptation to SR. Various SSL methods for SR, proposed in the literature and built upon these frameworks, are also presented. An extensive review of these approaches is then conducted: (1) the effect of the main hyperparameters of SSL frameworks is investigated; (2) the role of SSL components is studied (e.g., data-augmentation, projector, positive sampling); and (3) SSL frameworks are evaluated on SR with in-domain and out-of-domain data, using a consistent experimental setup, and a comprehensive comparison of SSL methods from the literature is provided. Specifically, DINO achieves the best downstream performance and effectively models intra-speaker variability, although it is highly sensitive to hyperparameters and training conditions, while SimCLR and MoCo provide robust alternatives that effectively capture inter-speaker variability and are less prone to collapse. This work aims to highlight recent trends and advancements, identifying current challenges in the field.
\end{abstract}




\end{frontmatter}



\section{Introduction}

Speaker Recognition (SR) refers to the task of determining a person's identity based on voice characteristics, making it a fundamental aspect of speech processing. It encompasses two main objectives: Speaker Identification (SID), which assigns an unknown voice to one of several enrolled speakers, and Speaker Verification (SV), which determines whether an unknown voice matches a claimed identity. SR has numerous applications, including forensic analysis, access control, and security systems, where voice-based authentication offers an alternative to traditional biometric modalities, such as fingerprints or facial recognition \cite{bimbot2004TutorialSV}. This work focuses exclusively on SV, as it aligns with the primary applications of SR in real-world systems.

The common principle behind SV methods is to learn speech representations that maximize inter-speaker distances and minimize intra-speaker variances, while ensuring robustness to extrinsic variabilities, such as speech context and emotion, noise, environment, recording, and channel conditions. These systems have undergone significant evolution, transitioning from statistical modeling to deep learning approaches. Early methods relied on the i-vector~\cite{dehak2011IVector}, a compact representation derived from a probabilistic framework involving Gaussian Mixture Models (GMM) and Universal Background Models (UBM) \cite{reynolds2000SVGMM}. Along with the emergence of deep learning, the x-vector \cite{snyder2018XVectors} approach was proposed to learn speaker representations by training a Deep Neural Network (DNN) to match speech utterances to their correct speaker identities in a supervised manner. Traditional machine learning methods, which combined front-end feature extraction and back-end scoring systems, have been surpassed by end-to-end DNN-based models, where scoring is performed by computing the cosine similarity between speaker representations \cite{bai2021SpeakerRecognitionDLOverview}. Other model architectures, such as VGG \cite{nagrani2017VoxCeleb}, ResNet \cite{chung2020DelvingVoxCeleb}, ECAPA-TDNN \cite{desplanques2020ECAPATDNN}, Res2Net~\cite{zhou2021Res2NetSV}, and MFA-Conformer \cite{zhang2022MFAConformer}, have also been applied, and subsequent advancements have refined these methods by exploring improved pooling strategies and training objectives \cite{cai2018ExploringE2ESV,chung2020DefenceMetricLearningSR,villalba2020StateOfTheArtSRNeuralNetwork}. These developments have been driven by SR labeled datasets, such as the NIST SRE corpora \cite{doddington2000NISTSRE} and VoxCeleb \cite{nagrani2017VoxCeleb,chung2018VoxCeleb2}. However, as with many deep learning approaches, the performance of these systems scales with larger amounts of training data \cite{sun2017EffectivenessDataDeepLearning}. This reliance on vast labeled datasets poses a significant challenge, as annotated samples are often scarce and costly to obtain, limiting the scalability and accessibility of these methods.

Self-Supervised Learning (SSL) methods have been developed to overcome this limitation by enabling models to learn relevant representations directly from the input data without human supervision. SSL has demonstrated its ability to improve scalability by exploiting the vast amounts of unlabeled data that are readily available. The standard framework involves training models on a general \textit{pretext} task and applying the learned representations to a \textit{downstream} task for specific applications, with an optional supervised fine-tuning step. SSL demonstrated its potential in Natural Language Processing (NLP) with models such as BERT \cite{devlin2019BERT}, leveraging \textit{sequence prediction} pretext objectives, which consist of predicting masked parts of a sequence (\textit{masked language modeling}). The same strategy was widely adopted in Speech Processing (SP), particularly for Automatic Speech Recognition (ASR), leading to models such as wav2vec 2.0 \cite{baevski2020wav2vec2.0}, HuBERT \cite{hsu2021HuBERT}, UniSpeech \cite{wang2021UniSpeech}, data2vec \cite{baevski2022data2vec}, and WavLM \cite{chen2022WavLM}. These approaches represent milestones in self-supervised speech representation learning, as they demonstrated that models can learn meaningful acoustic and linguistic structures directly from raw waveforms. Early work such as Contrastive Predictive Coding (CPC) \cite{vandenoord2019CPC} and the original wav2vec model \cite{schneider2019wav2vec} laid the groundwork by introducing contrastive learning objectives to predict future latent representations. Subsequent frameworks extended this concept of predicting masked or missing information: wav2vec 2.0 \cite{baevski2020wav2vec2.0} employs a contrastive objective on masked latent representations, while HuBERT \cite{hsu2021HuBERT} predicts discrete cluster assignments for masked frames. More recent models, such as WavLM \cite{chen2022WavLM}, further advance this paradigm by integrating robust pretraining strategies that improve generalization across diverse speech domains and noisy conditions. This line of research established self-supervised pretraining as a cornerstone of modern ASR systems. In Computer Vision (CV), SSL primarily relies on \textit{instance-invariance} pretext objectives, which involve maximizing the similarity between representations of the same class, encompassing various paradigms such as \textit{contrastive learning}, \textit{clustering}, \textit{information maximization}, and \textit{self-distillation}. SimCLR \cite{chen2020SimCLR}, MoCo \cite{he2020MoCo}, and DINO \cite{caron2021DINO} are some examples of these frameworks and will be covered in further detail in the following.

SSL has also been successfully applied to SV, with two main strategies emerging: (1) fine-tuning a large-scale SSL foundation model, pre-trained on speech data, with speaker identity labels \cite{fan2021ExploringWav2vec2.0SV,chen2022WhySSLASRBenefitSR,chen2022SSLASRforSV,novoselov2022SRwithWav2vec2.0} and (2) training an SSL model on unlabeled data and directly using the learned representations on SV. Although the former achieves state-of-the-art performance, it is not the focus of this work because it relies on annotated samples. The latter approach is based on SSL \textit{instance-invariance} frameworks, drawing parallels to CV, where similar methods are used for downstream tasks related to image processing. Note that SV does not require fine-tuning the SSL model, unlike many downstream tasks (e.g., classification, segmentation, transcription), since verification scores can be inferred directly from the representations due to the similarity between the pretext and downstream tasks.

\begin{figure*}
  \centering
  \subfloat[\textbf{Training} \textit{(Pretext task)}\label{fig:SSL_SV_framework:train}]{
    \includegraphics[width=0.71\linewidth]{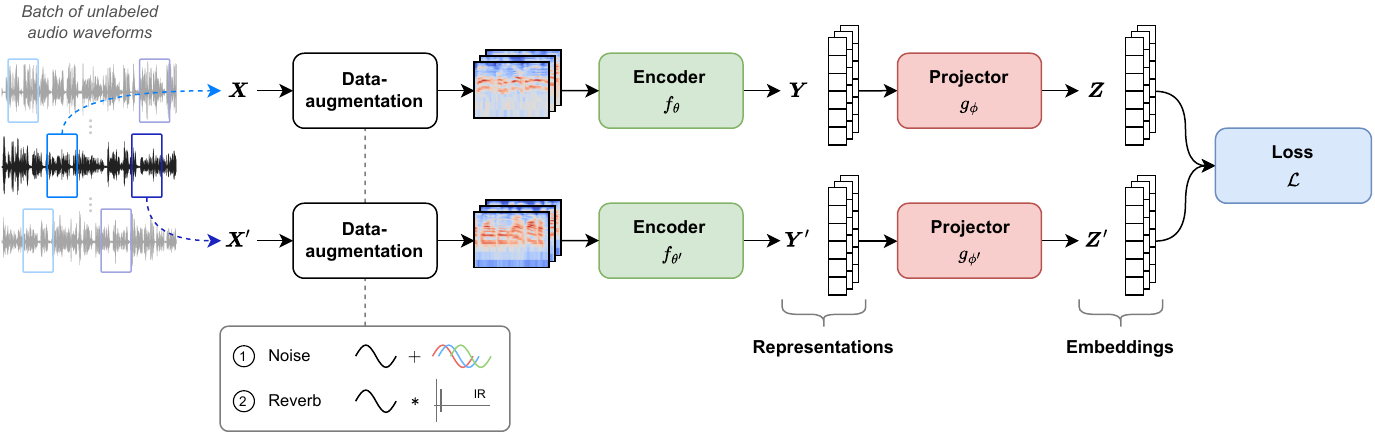}
  }
  \subfloat[\textbf{Evaluation} \textit{(Downstream task)}\label{fig:SSL_SV_framework:eval}]{
    \includegraphics[width=0.28\linewidth]{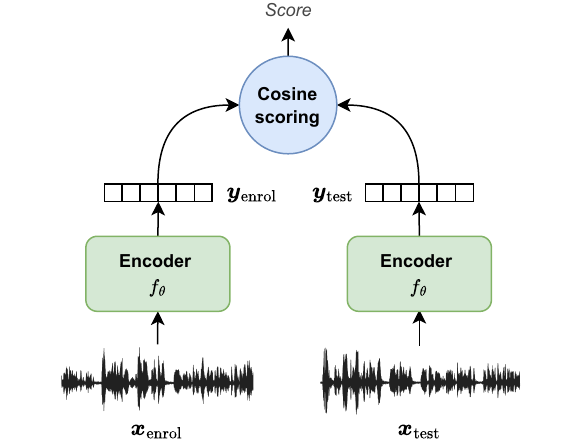}
  }
  \titlecaption{\acf{SSL} framework for \acf{SV}}{The training \textbf{(a)} is performed on a \textit{pretext} task to learn relevant representations, which will be used to perform the evaluation \textbf{(b)} on the \textit{downstream} task. The training framework adopts the \textit{joint-embedding} architecture to generate a pair of embeddings (\textbf{\textcolor[HTML]{007FFF}{\textit{anchor}}} and \textbf{\textcolor[HTML]{1B25BD}{\textit{positive}}}) from an unlabeled audio waveform.}
  \label{fig:SSL_SV_framework}
\end{figure*}

SSL frameworks for \textit{instance-invariance} are based on the \textit{joint-embedding} architecture (Figure~\subref*{fig:SSL_SV_framework:train}), generating representations from an \textit{anchor} and a \textit{positive} sample derived from distinct augmented views of the same input, thereby representing the same high-level attributes \cite{balestriero2023CookbookSSL}. These methods aim to maximize the similarity of \textit{anchor}-\textit{positive} pairs and employ different mechanisms to prevent a collapsing solution (i.e., a trivial solution leading to non-informative representations). In the context of SV, the fundamental assumption is that the anchor and positive originate from the same speaker identity, since both are extracted from the same speech utterance. Therefore, providing different versions of the same input audio with data-augmentation is fundamental for learning robust representations that primarily capture speaker identity rather than extrinsic information (e.g., noise conditions, room acoustics, microphone characteristics, transmission system, compression codecs) invariant within utterances \cite{zhang2021SimCLR,xia2021MoCo,lepage2022LabelEfficient}.

Contrastive learning frameworks, such as CPC \cite{vandenoord2019CPC}, SimCLR \cite{chen2020SimCLR}, and MoCo \cite{he2020MoCo}, aim to maximize the similarity of anchor-positive pairs while minimizing the similarity of anchor-negative pairs, where negatives are sampled from all other utterances within the current training batch or a memory queue. Clustering frameworks, like DeepCluster \cite{caron2018DeepCluster} and SwAV \cite{caron2020SwAV}, learn by grouping representations into clusters and using the resulting assignments as supervisory signals. Information maximization frameworks, such as W-MSE \cite{ermolov2021WMSE}, Barlow Twins \cite{zbontar2021BarlowTwins}, and VICReg \cite{bardes2022VICReg}, impose statistical constraints on the embeddings to ensure invariance between anchor and positive embeddings while explicitly preventing collapse. Self-distillation frameworks, like BYOL \cite{grill2020BYOL}, SimSiam \cite{chen2021SimSiam}, and DINO \cite{caron2021DINO}, are based on knowledge distillation and the student-teacher paradigm, where the \textit{student} model learns to predict the output of the \textit{teacher} model, while the latter is bootstrapped from the former and iteratively refined.

Based on these SSL frameworks, various methods have been developed specifically for SV, which can be categorized into \textit{single-stage} and \textit{multi-stage} approaches. Single-stage methods correspond to models trained with an SSL objective. Multi-stage methods correspond to models trained using pseudo-labels generated from a single-stage SSL model in a supervised fashion. Most of the methods are based on \textit{contrastive learning} \cite{huh2020AAT,zhang2021SimCLR,xia2021MoCo,lepage2024AdditiveMargin} and \textit{self-distillation} \cite{zhang2022C3DINO,chen2023RDINO,han2024CADINO,cho2022DINO,heo2022DINOCurriculum} since DINO is currently at the core of state-of-the-art approaches. The focus of single-stage methods has been to complement traditional SSL frameworks through various aspects, such as reducing the encoding of channel information \cite{huh2020AAT,kang2022LMix}, developing new positive sampling strategies \cite{tao2023DPP,han2024CADINO,lepage2025SSPS}, improving speaker separability \cite{lepage2024AdditiveMargin}, addressing class collisions \cite{xia2021MoCo}, and evaluating the complementarity of different frameworks \cite{lepage2022LabelEfficient,sang2022SSReg,chen2023RDINO}. Most multi-stage methods have been proposed for the SSL tracks of the VoxCeleb Speaker Recognition Challenge (VoxSRC) \cite{huh2024VoxSRC} and aim to mitigate label noise from unreliable pseudo-labels \cite{wang2020DKUDukeECEVoxSRC20,thienpondt2020IDLABVoxSRC20,tao2022LGL,han2022DLGLC} or rely on multi-modal audio-visual representations \cite{cai2021DKUDukeECEVoxSRC21,tao2023DPP,han2024CADINO} to further enhance downstream performance.

Given the growing interest in SSL, several reviews have been published covering the technology \cite{ericsson2022SelfSupervisedRepresentationLearning,balestriero2023CookbookSSL,gui2024SSLSurvey}, as well as its applications in NLP \cite{liu2023SSLSurveyNLP,xia2020BERTSurvey,qiu2020SSLSurveyNLP}, CV \cite{jing2021SSLSurveyCV}, and SP \cite{borgholt2022SSLSurveySpeech,mohamed2022SSLSurveySpeech}. A brief overview of SSL \textit{self-distillation} methods for SV has been published \cite{chen2022ComprehensiveStudySelfDistillation}. However, none of these works provides a dedicated and extensive analysis of SSL for SV. Several aspects of SSL remain unclear in this context, including its underlying mechanisms, key properties, and limitations. The community has developed numerous methods, yet no clear direction or consensus has emerged regarding the best approaches to consider. Therefore, this work presents the first review and study of the application of SSL to the downstream task of SV.

The contributions of this work are outlined below:
\begin{itemize}
    \item A comprehensive comparison of all major SSL frameworks on SV benchmarks, evaluated under consistent experimental conditions;
    \item An overview and comparison of single- and multi-stage SSL methods for SV from the literature;
    \item An analysis of the effect of key hyperparameters in SSL frameworks (e.g., number of negatives, temperatures, momentum update coefficient);
    \item A study of the role of various SSL components (e.g., data-augmentation, projector, positive sampling);
    \item The identification of the primary bottlenecks in current SSL approaches, with several suggestions for future research directions.
    \item \textit{sslsv}\footnote{The \textit{sslsv} toolkit used to conduct all experiments presented in this article is publicly available at \url{https://github.com/theolepage/sslsv}.}: an open-source PyTorch-based toolkit for training and evaluating SSL frameworks on SV, which enables the reproduction of all results presented in this article.
\end{itemize}

The content of this article is organized as follows. First, major SSL frameworks are presented and applied to SV (Section~\ref{sec:ssl}). Then, an overview of single- and multi-stage SSL methods for SV, proposed in the literature, is provided (Section~\ref{sec:sslsv}). Subsequently, the experimental setup is described (Section~\ref{sec:exp_setup}), followed by an investigation into the role of SSL hyperparameters (Section~\ref{sec:hyperparams}) and a detailed study of the core components of SSL frameworks (Section~\ref{sec:study}). Then, SSL frameworks are evaluated and various SSL methods for SV are compared on VoxCeleb benchmarks (Section~\ref{sec:evaluation}). Finally, the article concludes and discusses future research directions (Section~\ref{sec:conclusions}).

\newcommand{\I}{\mathcal{I}}
\newcommand{\B}{\mathcal{B}}

\RenewDocumentCommand{\L}{o o}{%
  \mathcal{L}%
  \IfValueT{#1}{_{\text{#1}}}%
  \IfValueT{#2}{^{\text{(#2)}}}%
}

\renewcommand{\xi}{\boldsymbol{x}_i}
\newcommand{\xip}{\boldsymbol{x}_i^{\prime}}
\newcommand{\yi}{\boldsymbol{y}_i}
\newcommand{\yip}{\boldsymbol{y}_i^{\prime}}
\newcommand{\zi}{\boldsymbol{z}_i}
\newcommand{\zip}{\boldsymbol{z}_i^{\prime}}
\renewcommand{\pi}{\boldsymbol{p}_i}
\newcommand{\pip}{\boldsymbol{p}_i^{\prime}}

\newcommand{\Drepr}{D_{\text{r}}}
\newcommand{\Demb}{D_{\text{e}}}

\newcommand{\X}{\boldsymbol{X}}
\newcommand{\Xp}{\boldsymbol{X}^{\prime}}
\newcommand{\Y}{\boldsymbol{Y}}
\newcommand{\Yp}{\boldsymbol{Y}^{\prime}}
\newcommand{\Z}{\boldsymbol{Z}}
\newcommand{\Zp}{\boldsymbol{Z}^{\prime}}

\newcommand{\f}{f}
\newcommand{\ftheta}{f_{\theta}}
\newcommand{\fthetap}{f_{\theta^{\prime}}}
\renewcommand{\g}{g}
\newcommand{\gphi}{g_{\phi}}
\newcommand{\gphip}{g_{\phi^{\prime}}}
\newcommand{\h}{h}
\newcommand{\hpsi}{h_{\psi}}
\newcommand{\hpsip}{h_{\psi^{\prime}}}

\section{Self-Supervised Learning Frameworks}
\label{sec:ssl}

The SSL training framework adopts the \textit{joint-embedding} architecture, depicted in Figure~\subref*{fig:SSL_SV_framework:train}, to generate a pair of \textit{representations} and a pair of \textit{embeddings} from an unlabeled audio waveform. The objective is to maximize the similarity of different views of the same input in latent space, by assuming that each training sample contains speech from a single speaker.

Let $\I \equiv\{1, \ldots, N\}$ be the indices of the training set of $N$ samples. At each iteration, $B$ utterances are sampled to create a batch with indices $\B \subseteq \I$. From each utterance $u_i$ with $i \in \B$, two segments, denoted by $\xi$ (\textit{anchor}) and $\xip$ (\textit{positive}), are randomly extracted. Random data-augmentation preprocessing is applied to both segments, and their mel-scaled spectrogram features are used as input for the model. Applying data-augmentation to generate different views of the same input audio, commonly through background noise addition and reverberation, is essential for emphasizing speaker identity over extrinsic factors, which are invariant within utterances.

The joint-embedding architecture comprises two branches: (1) encoder $\ftheta$ and projector $\gphi$; (2) encoder $\fthetap$ and projector $\gphip$. $\f$ and $\g$ share the same architecture across branches. The encoders $\ftheta$ and $\fthetap$ create representations $\yi=\ftheta(\xi)$ and $\yip=\fthetap(\xip)$ of dimension $\Drepr$. The projectors $\gphi$ and $\gphip$ create embeddings $\zi=\gphi(\yi)$ and $\zip=\gphip(\yip)$ of dimension $\Demb$. The \textit{representations} are used for the SV downstream task (Figure~\subref*{fig:SSL_SV_framework:eval}), while the \textit{embeddings} are used to compute the objective function $\L$ during training (Figure~\subref*{fig:SSL_SV_framework:train}). The projector is discarded for some frameworks, since it degrades performance, as detailed in Section~\ref{sec:study:projector}, in which case $g$ is the identity function (i.e., embeddings are identical to representations).

The default training scheme adopts the \textit{symmetrical} version of the architecture where the weights are shared across branches: $\theta^{\prime} = \theta$ and $\phi^{\prime} = \phi$. The \textit{asymmetrical} variant (i.e., MoCo, BYOL, and DINO) dissociates the two branches, with the first referred to as \textit{student} or \textit{query}, and the second as \textit{teacher} or \textit{key}. In this case, no gradient is propagated through the second branch, via a stop-gradient (sg) operation, and the weights of the second branch are updated using an Exponential Moving Average (EMA) of the weights of the first branch: $\theta^{\prime} \gets m\theta^{\prime} + (1-m)\theta$ and $\phi^{\prime} \gets m\phi^{\prime} + (1-m)\phi$, with $m \in [0, 1)$ the momentum update factor.

As batches of utterances are processed during training, the following notations are introduced: $\X=\{\xi\}_{i \in \B}$, $\Xp=\{\xip\}_{i \in \B}$, $\Y=\{\yi\}_{i \in \B}$, $\Yp=\{\yip\}_{i \in \B}$, $\Z=\{\zi\}_{i \in \B}$, and $\Zp=\{\zip\}_{i \in \B}$.

In the following, $\zi$ is the $i$-th embedding of $\Z$, $\boldsymbol{z}^d$ is the $d$-th dimension of embeddings in $\Z$, and $\ell(\boldsymbol{a}, \boldsymbol{b})= \exp\left(\cossim\left(\boldsymbol{a}, \boldsymbol{b}\right) / \tau\right)$ where $\tau$ is a temperature hyperparameter and $\cossim(\boldsymbol{a}, \boldsymbol{b})$ is the cosine similarity between the vectors $\boldsymbol{a}$ and $\boldsymbol{b}$.

\begin{figure*}
  \newlength{\imgwidth}
  \setlength{\imgwidth}{0.18\linewidth}
  \centering
  \subfloat[\textbf{Contrastive learning}\label{fig:SSL_conceptual_comparison:contrastive}]{
    \includegraphics[width=\imgwidth, trim=0.44cm 0 1.16cm 0, clip]{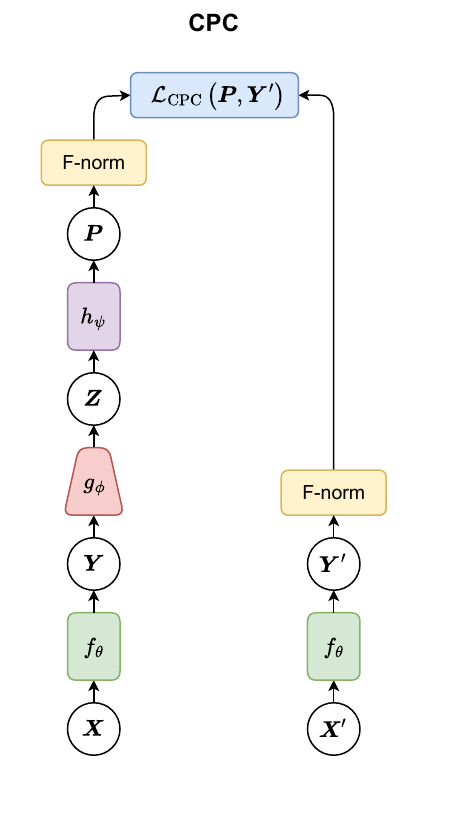}
    \includegraphics[width=\imgwidth, trim=0.44cm 0 1.16cm 0, clip]{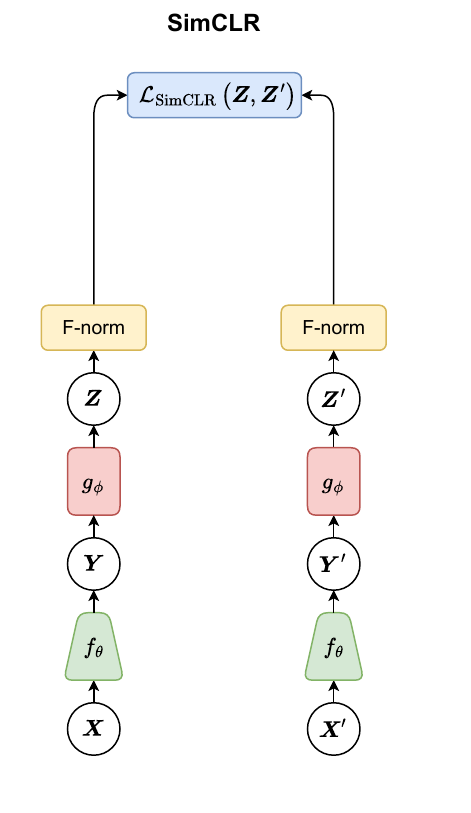}
    \includegraphics[width=\imgwidth, trim=0.44cm 0 1.16cm 0, clip]{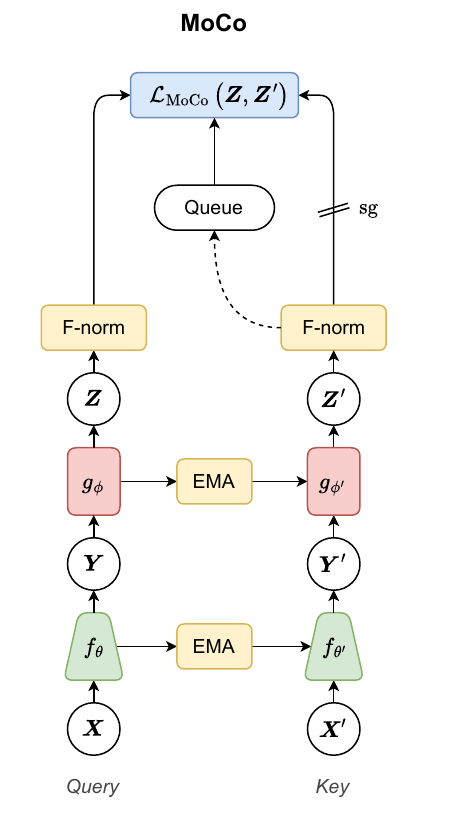}
  }%
  \hspace{0.05\linewidth}%
  \subfloat[\textbf{Clustering}\label{fig:SSL_conceptual_comparison:clustering}]{
    \includegraphics[width=\imgwidth, trim=0.44cm 0 1.16cm 0, clip]{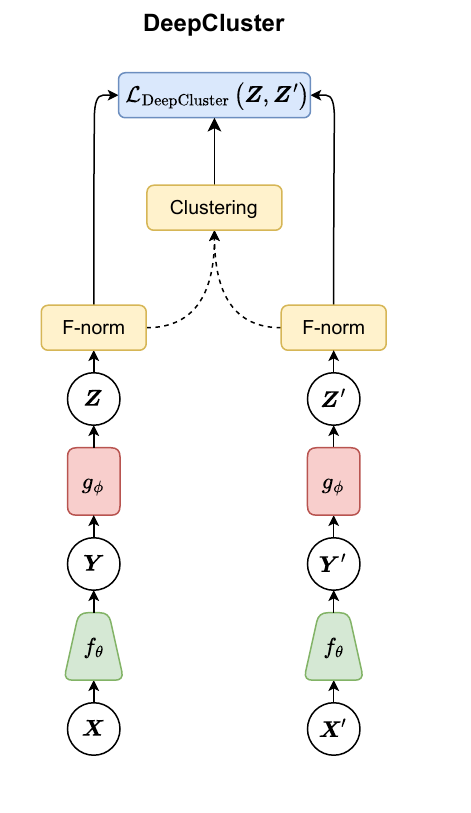}
    \includegraphics[width=\imgwidth, trim=0.44cm 0 1.16cm 0, clip]{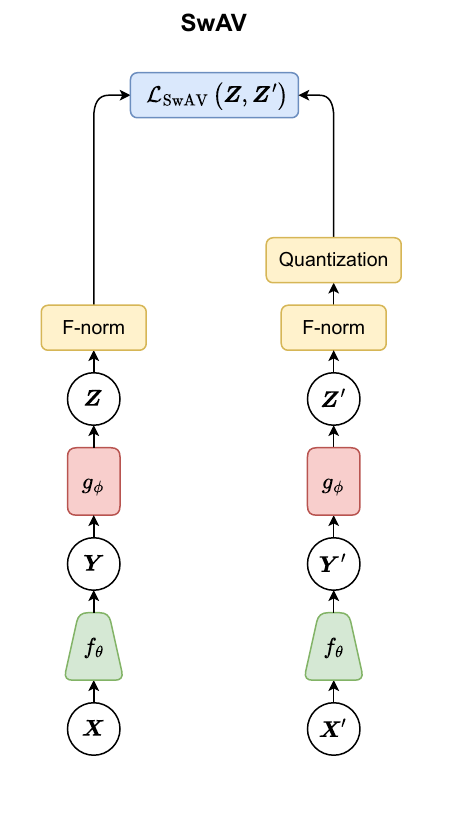}
  }
  \vspace{0.3cm}
  \subfloat[\textbf{Information maximization}\label{fig:SSL_conceptual_comparison:infomax}]{
    \includegraphics[width=\imgwidth, trim=0.44cm 0 1.16cm 0, clip]{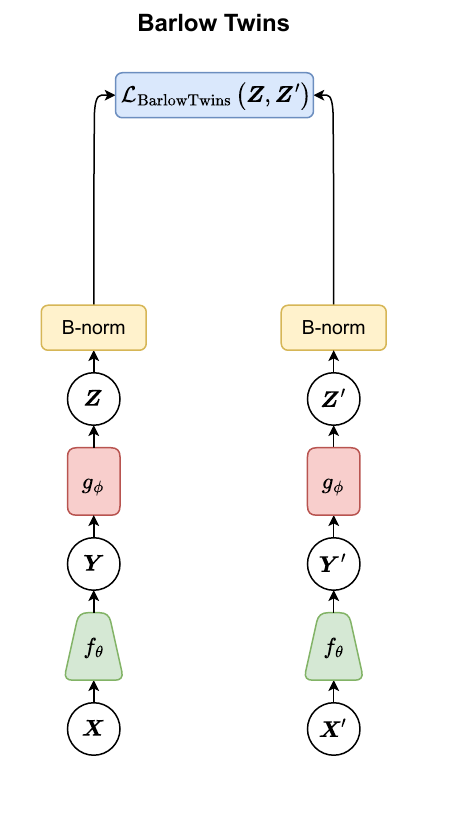}
    \includegraphics[width=\imgwidth, trim=0.44cm 0 1.16cm 0, clip]{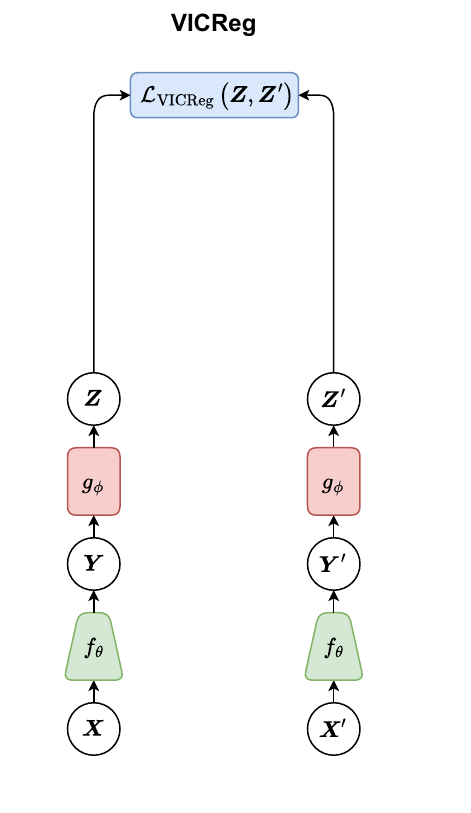}
  }%
  \hspace{0.05\linewidth}%
  \subfloat[\textbf{Self-distillation}\label{fig:SSL_conceptual_comparison:distillation}]{
    \includegraphics[width=\imgwidth, trim=0.44cm 0 1.16cm 0, clip]{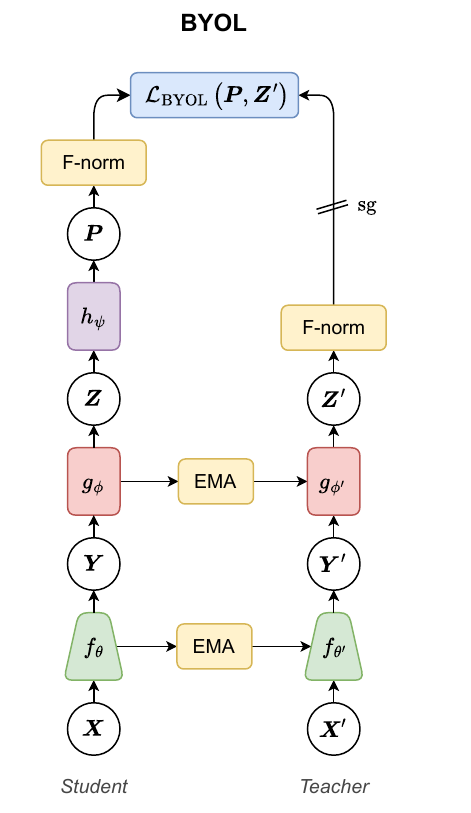}
    \includegraphics[width=\imgwidth, trim=0.44cm 0 1.16cm 0, clip]{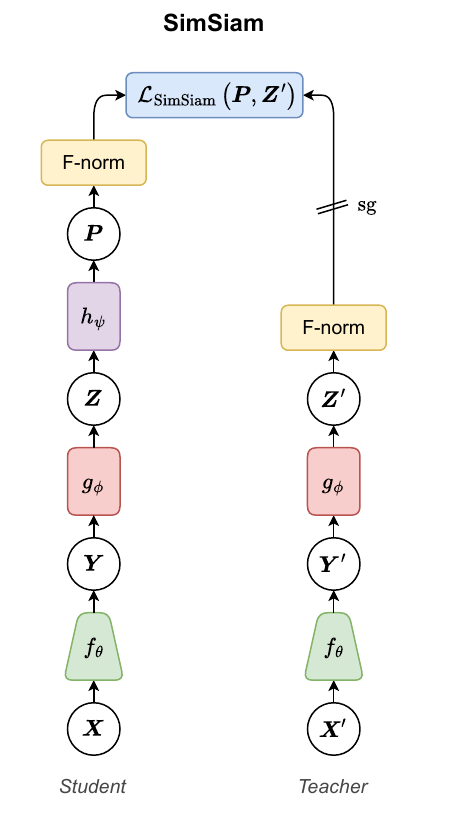}
    \includegraphics[width=\imgwidth, trim=0.44cm 0 1.16cm 0, clip]{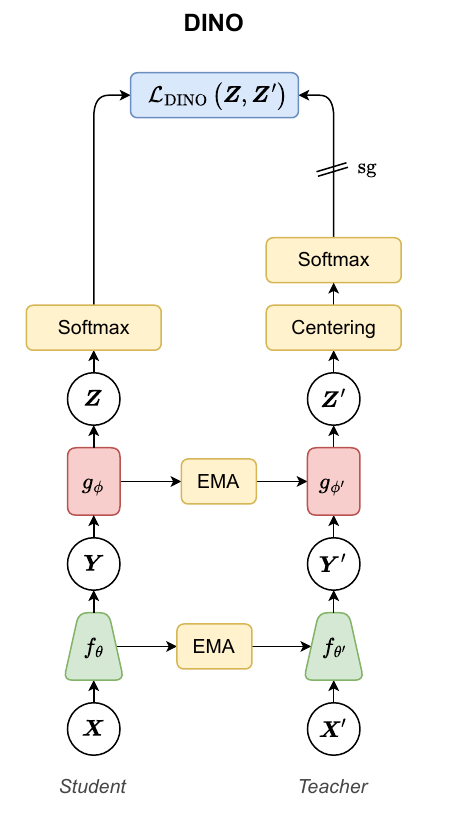}
  }
  \titlecaption{Conceptual comparison of different \ac{SSL} frameworks across various paradigms}{
    The \textit{joint-embedding} architecture remains constant, but the mechanisms to prevent collapse vary between methods. Encoders are represented in \textbf{\textcolor[HTML]{82B366}{green}}, projectors in \textbf{\textcolor[HTML]{B85450}{red}}, predictors in \textbf{\textcolor[HTML]{9673A6}{purple}}, operations in \textbf{\textcolor[HTML]{D6B656}{yellow}}, and losses in \textbf{\textcolor[HTML]{6C8EBF}{blue}}. Rectangular-shaped modules process fixed-length feature vectors or sequences of feature vectors, while trapezoid-shaped modules, functioning as encoders or autoregressive models, reduce variable-length sequences into fixed-length feature vectors. \texttt{F-norm} refers to feature normalization, \texttt{B-norm} to batch normalization, \texttt{EMA} to Exponential Moving Average, and \texttt{sg} to stop-gradient.
  }
  \label{fig:SSL_conceptual_comparison}
\end{figure*}

This section presents the main SSL \textit{instance-invariance} frameworks from different paradigms, which have initially been proposed for CV downstream tasks: \textit{contrastive learning} (Section~\ref{sec:ssl:contrastive}), \textit{clustering} (Section~\ref{sec:ssl:clustering}), \textit{information maximization} (Section~\ref{sec:ssl:infomax}), and \textit{self-distillation} (Section~\ref{sec:ssl:distillation}). These frameworks aim to maximize the similarity of \textit{anchor}-\textit{positive} pairs, representing the same speaker identity, and rely on different techniques to prevent collapse. A conceptual comparison of these approaches is provided in Figure~\ref{fig:SSL_conceptual_comparison}.

\subsection{Contrastive learning}
\label{sec:ssl:contrastive}

Contrastive learning frameworks (Figure~\subref*{fig:SSL_conceptual_comparison:contrastive}) aim at maximizing the similarity of anchor-positive pairs while minimizing the similarity of anchor-negative pairs, where negatives are sampled from the training batch or a larger memory queue by assuming that they belong to a distinct speaker identity. The impact of ``class collisions'' (i.e., false negatives arising from random negative sampling) is generally considered negligible. The existence of negatives pairs is the main component preventing collapse, as demonstrated in Section~\ref{sec:study:collapse}.

\subsubsection{\textbf{CPC}}
\label{sec:ssl:contrastive:cpc}
CPC (\textit{Contrastive Predictive Coding}) \cite{vandenoord2019CPC, henaff2020CPCV2} learns representations by predicting future frames in latent space using aggregated information from past frames. The contrastive loss induces the latent space to capture the most valuable information for predicting future timesteps. Unlike all other frameworks, CPC processes frame-level representations of utterances. Let $T$ be the total number of timesteps and $t_0$ the timestep from which predictions will be made. The encoder outputs representations $\{ \boldsymbol{y}^{(t)}_i \}_{1\leq t \leq T}$ from $\xi$ and $\{ \boldsymbol{y}^{\prime(t)}_i \}_{1\leq t \leq T}$ from $\boldsymbol{x}^{\prime}_i$. An autoregressive model $g_{\phi}$ transforms $\{ \boldsymbol{y}^{(t)}_i \}_{1\leq t \leq t_0}$ to $\boldsymbol{z}^{(t)}$ and a predictor model $\hpsi$ maps $\boldsymbol{z}^{(t)}$ to predictions $\{ \pi^{(t)} \}_{t_0+1\leq t \leq T}$. The $\L[CPC]$ loss is defined as:
\begin{equation}
    \label{eq:cpc_loss}
    \L[CPC] = - \frac{1}{B \cdot T} \sum_{i \in \B} \sum_{t=t_0+1}^{T} \log \frac{\ell \left(\pi^{(t)}, \boldsymbol{y}_i^{\prime(t)} \right)}{\sum\limits_{j \in \B} \ell \left(\pi^{(t)}, \boldsymbol{y}_j^{\prime(t)} \right)}.
\end{equation}

\subsubsection{\textbf{SimCLR}}
\label{sec:ssl:contrastive:simclr}
SimCLR \cite{chen2020SimCLR, chen2020SimCLRV2} is a simple contrastive framework where negatives are sampled from the current batch. Each anchor-positive pair is associated with $B-1$ anchor-negative pairs. The $\L[SimCLR]$ loss, referred to as the NT-Xent (Normalized Temperature-scaled Cross Entropy) objective function in the original paper \cite{chen2020SimCLR}, is defined as:
\begin{equation}
    \label{eq:simclr_loss}
    \L[SimCLR] = - \frac{1}{B} \sum_{i \in \B} \log \frac{\ell \left(\zi, \zip \right)}{\sum\limits_{j \in \B} \ell \left(\zi, \boldsymbol{z}_j^{\prime} \right)}.
\end{equation}
Note that the implementation is symmetric, resulting in $2B$ positive pairs with $2(B-1)$ negative pairs each.

\subsubsection{\textbf{MoCo}}
\label{sec:ssl:contrastive:moco}
MoCo \cite{he2020MoCo, chen2020MoCoV2} is a contrastive learning approach employing the \textit{asymmetrical} joint-embedding architecture and a large memory queue $\boldsymbol{Q}^{\text{MoCo}}$ of embeddings from previous iterations to increase the number of negative pairs further. The queue is updated at each training step with embeddings from the key branch. Each anchor-positive pair is associated with $|\boldsymbol{Q}^{\text{MoCo}}|$ negatives such that $|\boldsymbol{Q}^{\text{MoCo}}| \gg B$. The $\L[MoCo]$ loss is defined as:
\begin{equation}
    \label{eq:moco_loss}
    \L[MoCo] = - \frac{1}{B} \\ \sum_{i \in \B} \log \frac{\ell \left(\zi, \zip\right)}{\ell \left(\zi , \zip \right) + \sum\limits_{j \in \mathcal{J}} \ell \left(\zi, \boldsymbol{q}_j \right)},
\end{equation}
where $\mathcal{J} \equiv\{1, \ldots, |\boldsymbol{Q}^{\text{MoCo}}|\}$ are the memory queue indices and $\boldsymbol{q}_j$ is the $j$-th element from the queue.

\subsection{Clustering}
\label{sec:ssl:clustering}

Clustering frameworks (Figure~\subref*{fig:SSL_conceptual_comparison:clustering}) generate assignments from representations by grouping them into discrete clusters, enabling the model to learn discriminative features by predicting these cluster assignments.

\subsubsection{\textbf{DeepCluster}}
\label{sec:ssl:clustering:deepcluster}
DeepCluster \cite{caron2018DeepCluster} iteratively groups all training embeddings learned via SSL with a clustering algorithm and uses the assignments as supervision to optimize the model. Clustering is performed using k-means at the beginning of each epoch on all embeddings from the previous epoch. The resulting labels are used to optimize a standard cross-entropy classification loss. The $\L[DeepCluster]$ loss is defined as:
\begin{equation}
    \label{eq:deepcluster_loss}
    \L[DeepCluster] = - \frac{1}{B} \sum_{i \in \B} \log \frac{\ell \left(\zi, \boldsymbol{c}_{c_i}\right)}{\sum\limits_{k=1}^{K} \ell \left(\zi , \boldsymbol{c}_k \right)},
\end{equation}
where $K$ is the number of clusters, $\boldsymbol{c}_k$ is the prototype (centroid) of the $k$-th cluster and $c_i$ is the assigned cluster for the $i$-th sample.
Note that the implementation is symmetric (i.e., the classification is performed on both anchor and positive embeddings, which is equivalent to the single-branch method described in the original paper \cite{caron2018DeepCluster}), and the clustering is repeated multiple times to optimize the model with different sets of pseudo-labels from different views.

\subsubsection{\textbf{SwAV}}
\label{sec:ssl:clustering:swav}
SwAV \cite{caron2020SwAV} learns to map embeddings to a set of clusters in an online fashion. Prototypes are jointly optimized with the model, and soft assignments are created using the Sinkhorn-Knopp algorithm \cite{cuturi2013Sinkhorn}. Sinkhorn-Knopp balances two objectives: (1) maximizing the similarity between embeddings and centroids and (2) ensuring balanced cluster sizes by uniformly distributing the assignments. The ``swapped" mechanism, where the code of the anchor is predicted from the positive embedding and vice versa, enforces consistency between assignments produced from different views of the same utterance. Note that a queue $\boldsymbol{Q}^{\text{SwAV}}$ of latest embeddings is introduced after several epochs to provide more stable assignments. The $\L[SwAV]$ loss is defined as:
\begin{equation}
    \label{eq:swav_loss}
    \L[SwAV] = - \frac{1}{B \cdot P} \sum_{i \in \B} \sum_{k=1}^{P} \boldsymbol{a}_{i,k} \log \frac{\ell \left(\zip, \boldsymbol{c}_{k}\right)}{\sum\limits_{k^{\prime}=1}^{P} \ell \left(\zip , \boldsymbol{c}_{k^{\prime}} \right)},
\end{equation}
where $P$ is the number of prototypes, $\boldsymbol{a}_{i, k}$ is the assignment for the $i$-th anchor embedding and $k$-th prototype, and $\boldsymbol{c}_k$ is the $k$-th prototype. Note that the swapped mechanism is not represented in the objective function for conciseness.

\subsection{Information maximization}
\label{sec:ssl:infomax}

Information maximization frameworks (Figure~\subref*{fig:SSL_conceptual_comparison:infomax}) learn meaningful representations by maximizing the similarity between embeddings of anchor-positive pairs while minimizing redundancy across their dimensions to avoid collapse.

\subsubsection{\textbf{W-MSE}}
\label{sec:ssl:infomax:wmse}
W-MSE \cite{ermolov2021WMSE} maximizes the similarity of anchor-positive pairs while ensuring that the embeddings lie within a spherical distribution using a whitening projection. The whitening operation transforms the embeddings into the eigenspace of their covariance matrix, which has a ``scattering'' effect on the batch samples, preventing degenerate solutions. The $\L[W-MSE]$ loss is defined as:
\begin{equation}
    \label{eq:wmse_loss}
    \L[W-MSE] = 2 - 2 \cdot \frac{1}{B} \sum_{i \in \B} \cossim \left( W(\zi), W(\zip) \right),
\end{equation}
where $W(\boldsymbol{z})= \boldsymbol{W} (\boldsymbol{z}-\boldsymbol{\mu})$ corresponds to the whitening projection, $\boldsymbol{W}$ is the whitening matrix, and $\boldsymbol{\mu}$ is the mean of the embeddings. The whitening operation may be repeated over multiple iterations. For more details on how $\boldsymbol{W}$ is computed, refer to the original implementation \cite{ermolov2021WMSE}. Batch slicing is applied to mitigate the high variance that can arise when estimating the whitening matrix.

\subsubsection{\textbf{Barlow Twins}}
\label{sec:ssl:infomax:barlowtwins}
Barlow Twins \cite{zbontar2021BarlowTwins} maximizes the similarity of anchor-positive pairs by introducing constraints on the cross-correlation matrix $\mathcal{C}$ between $\Z$ and $\Zp$. The \textit{invariance} component forces the embeddings to be invariant to different transformations by pushing the on-diagonal coefficients to one. The \textit{redundancy reduction} component decorrelates the dimensions of the embeddings by pushing off-diagonal coefficients to zero. The complete objective function makes $\mathcal{C}$ similar to the identity matrix. The $\L[BarlowTwins]$ loss is defined as:
\begin{align}
    \label{eq:barlowtwins_loss}
    \L[BarlowTwins]
    &= \sum_{d=1}^{D}\left(1-\left[\mathcal{C}\left(\Z, \Zp\right)\right]_{d,d}\right)^{2} \notag \\
    & + \lambda \sum_{d=1}^{D} \sum_{\substack{d^{\prime}=1 \\ d^{\prime} \neq d}}^{D} \left[\mathcal{C}\left(\Z, \Zp\right)\right]_{d,d^{\prime}}^{2},
\end{align}
where $\lambda$ is a hyperparameter to scale the \textit{redundancy reduction} component.

\subsubsection{\textbf{VICReg}}
\label{sec:ssl:infomax:vicreg}
VICReg \cite{bardes2022VICReg} maximizes the similarity of anchor-positive pairs using multiple training objectives that provide an explicit solution to prevent collapse.
\begin{itemize}
    \item The \textit{variance} ($v$) term generates class-dependent representations, assuming that each sample in the batch is from a distinct speaker, and avoids collapse by enforcing the variance to reach one along the dimensions of the embeddings: $v\left(\Z\right)=\frac{1}{\Demb} \sum_{d=1}^{\Demb} \max \left(0, 1-\sqrt{\operatorname{Var}(\boldsymbol{z}^d)}\right)$.
    \item The \textit{invariance} ($s$) term learns invariance to different transformations of the same utterance by minimizing the Euclidean distance between anchor-positive embeddings: $s\left(\Z, \Zp\right)=\frac{1}{B} \sum_{i \in \B}\left\|\zi-\zip\right\|_{2}^{2}$.
    \item The \textit{covariance} ($c$) term prevents feature dimensions from encoding similar information by pushing off-diagonal coefficients of their covariance matrix $C$ to zero:\\ $c\left(\Z\right)=\frac{1}{\Demb} \sum_{d=1}^{\Demb} \sum_{\substack{d^{\prime}=1 \\ d^{\prime} \neq d}}^{\Demb} [C(\Z)]_{d, d^{\prime}}^{2}$.
\end{itemize}
The $\L[VICReg]$ loss is defined as:
\begin{align}
    \label{eq:vicreg_loss}
    \L[VICReg]
    &= \lambda \, s\left(\Z, \Zp\right) \notag \\
    &\quad + \mu\left(v(\Z) + v\left(\Zp\right)\right) \notag \\
    &\quad + \nu\left(c(\Z) + c\left(\Zp\right)\right),
\end{align}
where $\lambda$, $\mu$, and $\nu$ are hyperparameters to weight the \textit{invariance}, \textit{variance} and \textit{covariance} components.

\subsection{Self-distillation}
\label{sec:ssl:distillation}

Self-distillation frameworks (Figure~\subref*{fig:SSL_conceptual_comparison:distillation}) rely on the knowledge distillation training paradigm where a \textit{student} model learns to predict the output of a \textit{teacher} model. The intuition behind self-distillation is that the teacher will bootstrap supervision from the inherent properties of the data, even in the early stages of training, and that its capabilities will be progressively refined throughout the training. The main mechanisms to prevent collapse are the embedding normalizations and/or the asymmetry in the joint-embedding architecture, as shown in Section~\ref{sec:study:collapse}.

\subsubsection{\textbf{BYOL}}
\label{sec:ssl:distillation:byol}
BYOL \cite{grill2020BYOL} implements self-distillation and relies on the \textit{asymmetrical} joint-embedding architecture. A predictor module $\hpsi$ is introduced into the student branch to predict the output of the teacher projector. The $\L[BYOL]$ loss is defined as:
\begin{equation}
    \label{eq:byol_loss}
    \L[BYOL] = 2 - 2 \cdot \frac{1}{B} \sum_{i \in \B} \cossim \left( \pi, \zip \right),
\end{equation}
where $\pi = \hpsi(\zi)$ is the output of the predictor module. Note that the implementation is symmetric by recomputing the loss after swapping the anchor and positive inputs.

\subsubsection{\textbf{SimSiam}}
\label{sec:ssl:distillation:simsiam}
SimSiam \cite{chen2021SimSiam} is similar to BYOL except that the weights are shared between branches following the \textit{symmetrical} joint-embedding architecture. To avoid collapse, a stop-gradient operation is introduced when computing the loss to prevent the gradient from flowing through both branches. The $\L[SimSiam]$ loss is defined as:
\begin{equation}
    \label{eq:simsiam_loss}
    \L[SimSiam] = - \frac{1}{B} \sum_{i \in \B} \cossim \left( \pi, \text{sg}(\zip) \right),
\end{equation}
where $\pi = \hpsi(\zi)$ is the output of the predictor module and $\text{sg}$ is a stop-gradient operation. Note that the implementation is symmetric by recomputing the loss after swapping the anchor and positive inputs.

\subsubsection{\textbf{DINO}}
\label{sec:ssl:distillation:dino}
DINO \cite{caron2021DINO} is a self-distillation framework based on the \textit{asymmetrical} joint-embedding architecture. Following the ``multi-crop'' strategy \cite{caron2020SwAV}, a larger set of augmented utterances with different segment lengths is considered, resulting in $L$ short (\textit{local}) and $G$ long (\textit{global}) segments. Local and global views are fed through the student model, while only global views are fed through the teacher model to learn ``local-to-global'' relationships. The objective function optimizes the cross-entropy between the student and teacher output distributions. To prevent a collapsing solution, \textit{centering} and \textit{sharpening} are implemented for the teacher. The former avoids one dimension from prevailing, while the latter avoids collapse to the uniform distribution. The $\L[DINO]$ is defined as:
\begin{equation}
    \label{eq:dino_loss}
    \L[DINO] =
        \frac{1}{B} \sum_{i \in \B}
        \sum_{t=1}^{G} \sum_{\substack{s=1 \\ s \neq t}}^{G+L}
        H\bigg(
            \frac{\boldsymbol{z}_{i,t}^{\prime} - \boldsymbol{c}}{\tau_{\text{t}}},
            \frac{\boldsymbol{z}_{i,s}}{\tau_{\text{s}}}
        \bigg),
\end{equation}
where $H\left(\boldsymbol{a}, \boldsymbol{b}\right)=- \softmax(\boldsymbol{a}) \log \left( \softmax(\boldsymbol{b}) \right)$, $\boldsymbol{z}_{i,t}^{\prime}$ is the $t$-th teacher embedding of the $i$-th sample, $\boldsymbol{z}_{i,s}$ is the $s$-th student embedding of the $i$-th sample, $\tau_{\text{t}}$ is the teacher temperature, $\tau_{\text{s}}$ is the student temperature, and $\boldsymbol{c}$ is a running mean on the teacher outputs.

\begin{figure*}
  \centering
  \resizebox{\linewidth}{!}{\begin{tikzpicture}

    \definecolor{type1}{rgb}{0.2, 0.4, 0.8}
    \definecolor{type2}{rgb}{0.8, 0.4, 0.8}
    \definecolor{link}{rgb}{0.5, 0.5, 0.5}

    \def\dxNames{0.6}
    \def\dxDates{0.255}
    \def\yNames{2.5}
    \def\yDates{0}

    \newcommand{\addYear}[1]{ 
        \pgfmathtruncatemacro{\xDate}{#1 - 2020} 
        \fill (\xDate*12*\dxDates, \yDates) circle (1.5pt);
        \node[below, font=\scriptsize] at (\xDate*12*\dxDates, \yDates-0.1) {#1};
    }
    
    \newcommand{\addName}[4]{ 
        \def\xName{-0.8+#1*\dxNames}

        \pgfmathparse{#3} \let\fulldate\pgfmathresult 
        \pgfmathtruncatemacro{\year}{floor(\fulldate)} 
        \pgfmathtruncatemacro{\month}{round((\fulldate - \year) * 100)} 
        \pgfmathtruncatemacro{\pos}{(\year - 2020) * 12 + \month - 1}
        \pgfmathsetmacro{\xDate}{\pos * \dxDates}

        \draw[link, opacity=0.5, line width=0.2mm] (\xName, \yNames) .. controls (\xName, \yNames - 1) and (\xDate, \yDates + 1) .. (\xDate, \yDates);

        \fill[#4] (\xName, \yNames) circle (1.75pt);
        \node[above, anchor=south west, xshift=-5pt, yshift=1.5pt, rotate=0, font=\scriptsize] at (\xName, \yNames) {\rotatebox{30}{#2}};
        
        \fill[link] (\xDate, \yDates) circle (1.25pt);
    }

    \draw[black, -stealth, line width=0.2mm] (0, \yDates) -- (5*12*\dxDates + 10*\dxDates, \yDates);

    \foreach \label/\date/\colorname [count=\i from 0] in {
        {Disent. \tiny{\cite{nagrani2020Disentangled}}}/2020.05/type1,
        {CDDL \tiny{\cite{chung2020CDDL}}}/2020.10/type1,
        {CEL \tiny{\cite{mun2020CEL}}}/2020.10/type1,
        {GCL \tiny{\cite{inoue2020GCL}}}/2020.12/type1,
        {AAT \tiny{\cite{huh2020AAT}}}/2020.12/type1,
        {SimCLR + MSE loss \tiny{\cite{zhang2021SimCLR}}}/2021.06/type1,
        {MoCo + ProtoNCE \tiny{\cite{xia2021MoCo}}}/2021.06/type1,
        {SimCLR + AAT \tiny{\cite{tao2022LGL}}}/2022.05/type1,
        {SSReg \tiny{\cite{sang2022SSReg}}}/2022.05/type2,
        {i-mix / l-mix \tiny{\cite{kang2022LMix}}}/2022.08/type1,
        {DINO (Fast ResNet-34) \tiny{\cite{cho2022DINO}}}/2022.08/type2,
        {C3-MoCo \tiny{\cite{zhang2022C3DINO}}}/2022.08/type1,
        {C3-DINO \tiny{\cite{zhang2022C3DINO}}}/2022.08/type2,
        {DINO (RawNet3) \tiny{\cite{jung2022RawNet3}}}/2022.09/type2,
        {DINO + Cosine loss \tiny{\cite{han2022DLGLC}}}/2022.09/type2,
        {Contrastive + VICReg \tiny{\cite{lepage2022LabelEfficient}}}/2022.09/type1,
        {DINO + Aug. \tiny{\cite{chen2022ComprehensiveStudySelfDistillation}}}/2023.01/type2,
        {DPP \tiny{\cite{tao2023DPP}}}/2023.04/type1,
        {RDINO \tiny{\cite{chen2023RDINO}}}/2023.06/type2,
        {DINO + Curriculum \tiny{\cite{heo2022DINOCurriculum}}}/2023.08/type2,
        {CA-DINO \tiny{\cite{han2024CADINO}}}/2023.10/type2,
        {PDC-RDINO \tiny{\cite{zhao2024PrototypeDivisionDINO}}}/2024.03/type2,
        {W-ACSG \tiny{\cite{gan2024WACSG}}}/2024.04/type1,
        {SimCLR + Margin \tiny{\cite{lepage2024AdditiveMargin}}}/2024.06/type1,
        {MeMo \tiny{\cite{jin2024MeMo}}}/2024.09/type2,
        {RDINO + W-GVKT \tiny{\cite{jin2024WGVKT}}}/2024.09/type2,
        {DINO + RMP \tiny{\cite{kim2024RMP}}}/2024.09/type2,
        {SDPN \tiny{\cite{chen2025SDPN}}}/2025.04/type2,
        {EBCA-DINO \tiny{\cite{hao2025EBCADINO}}}/2025.04/type2,
        {SimCLR-SSPS \tiny{\cite{lepage2025SSPS}}}/2025.08/type1,
        {DINO-SSPS \tiny{\cite{lepage2025SSPS}}}/2025.08/type2%
    } {
        \addName{\i}{\label}{\date}{\colorname}
    }

    \foreach \year in {2020, 2021, 2022, 2023, 2024, 2025} {
        \addYear{\year}
    }

    \node[below, font=\footnotesize] at (8.625, -0.75) {
        \begin{tabular}{c@{\hskip 1cm}c}
            {\textcolor{type1}{\tikz\fill[scale=1pt] (0,0) circle (.5ex);}} Contrastive learning &
            {\textcolor{type2}{\tikz\fill[scale=1pt] (0,0) circle (.5ex);}} Self-distillation
        \end{tabular}
    };

\end{tikzpicture}}%
  \titlecaption{Timeline of a selection of single-stage \ac{SSL} methods for \ac{SV} from the literature}{
    Methods are categorized by framework: \textbf{\textcolor[rgb]{0.2, 0.4, 0.8}{Contrastive learning}} or \textbf{\textcolor[rgb]{0.8, 0.4, 0.8}{Self-distillation}}. The release date is determined according to the conference or journal publication date of the corresponding article.
  }
  \label{fig:timeline}
\end{figure*}

\section{Self-Supervised Methods for Speaker Verification}
\label{sec:sslsv}

Several methods have been proposed for the downstream task of SV. All the methods presented below are based on the SSL frameworks described in the previous section. Single-stage methods correspond to SSL models trained in an end-to-end fashion, while multi-stage methods correspond to supervised models trained using pseudo-labels generated from a single-stage SSL model and iteratively refined. This section lays the groundwork for comparing and evaluating these methods on SV in Section~\ref{sec:evaluation:methods}. An overview of the evolution of the literature on single-stage SSL for SV over the years is provided in Figure~\ref{fig:timeline}, highlighting the growing interest in this field and the gradual shift from contrastive learning to self-distillation approaches.

\subsection{Single-stage methods}

The most common single-stage methods are presented below, first grouped by their underlying framework, contrastive learning and self-distillation, and subsequently organized by theme within each framework.

\subsubsection{Contrastive learning}

\paragraph{General methods}
Given that contrastive learning represents the earliest self-supervised paradigm applied to SV, several methods have introduced broad extensions of contrastive frameworks without introducing any novel concepts.
\textbf{GCL} \textit{(Generalized Contrastive Loss)} \cite{inoue2020GCL} follows the SimCLR framework and unifies losses from supervised metric learning and unsupervised contrastive learning, enabling the application of the latter to semi-supervised scenarios (i.e., when the training set contains both labeled and unlabeled samples).

\paragraph{Cross-modal methods}
Cross-modal approaches leverage complementary information from multiple modalities, typically speech and visual data, to learn more robust speaker representations. These methods tend to enhance speaker discrimination but require additional modalities during training, which can limit their general applicability to audio-only training sets.
\textbf{Disent.} \textit{(Disentangled Speech Embeddings)} \cite{nagrani2020Disentangled} relies on the cross-modal relationships between speech and face representations from videos by performing video-to-audio matching within and across tracks while disentangling information between content and identity. \textbf{CDDL} \textit{(Cross-Domain Discriminative Loss)} \cite{chung2020CDDL} employs audio-to-video and video-to-audio matching objective functions and a cross-domain discriminative loss to enforce intra-class separation within modalities.

\paragraph{Regularization-based approaches}
To enhance the robustness of contrastive learning, regularization-based methods introduce additional loss terms to learn more relevant representations or stabilize training convergence. Such regularization can improve representation smoothness and generalization but often requires careful balancing with the main contrastive objective.
\textbf{CEL} \textit{(Contrastive Equilibrium Learning)} \cite{mun2020CEL} adopts the Angular Prototypical (AP) loss \cite{chung2020DefenceMetricLearningSR}, equivalent to the NT-Xent loss \cite{chen2020SimCLR} but with learnable scale and bias, and introduces a uniformity loss to further separate embeddings in the batch with a Gaussian pairwise similarity function. \textbf{SimCLR + MSE loss} \cite{zhang2021SimCLR} implements a simple regularization term to the SimCLR framework to prevent the model from encoding channel information by minimizing the Euclidean distance between pairs of clean and augmented embeddings of the same utterance. \textbf{Contrastive + VICReg} \cite{lepage2022LabelEfficient} explores the complementarity between the contrastive objective and the training components of VICReg (variance, invariance, and covariance) at different stages of the joint-embedding architecture.

\paragraph{Negative sampling and class-collision}
Among various extensions of contrastive learning, several methods focus on improving the negative sampling strategy and mitigating class collisions. By identifying false negatives or weighting difficult negatives, these approaches aim to enhance the reliability of contrastive pairs and the discriminative capacity of the resulting embeddings.
\textbf{C3-MoCo} \textit{(Class-Collision Correction)} \cite{zhang2022C3DINO}, equivalent to \textbf{MoCo + ProtoNCE} \cite{xia2021MoCo}, addresses the issue of class collisions, stemming from false-negative samples in the memory queue used by MoCo, by clustering speaker representations and relying on the ProtoNCE loss which considers prototypes from the same class as positives and prototypes from distinct classes as negatives. \textbf{W-ACSG} \textit{(Weighted Asymmetric Clean Segments-Guided)} \cite{gan2024WACSG} incorporates clean and augmented segments into the contrastive loss to generate additional positive and negative pairs and assigns larger weights to hard negative pairs based on their cosine similarity.

\paragraph{Mixup-based augmentation}
Mixup-based augmentation approaches generate new training samples by interpolating between existing ones to improve robustness and generalization.
\textbf{i-mix} / \textbf{l-mix} \cite{kang2022LMix} proposes to improve the generalization of speaker representations by creating diverse synthetic training samples following the mixup augmentation strategy \cite{zhang2018Mixup}: \textbf{i-mix} \textit{(instance mix)} performs interpolation on the input feature space, while \textbf{l-mix} \textit{(latent space i-mix)} performs interpolation on the representation space.

\paragraph{Margin-based discriminative learning}
By incorporating margins into the contrastive loss, either in cosine or angular space, margin-based methods aim to enhance inter-speaker separation and produce more discriminative and compact speaker representations.
\textbf{SimCLR + Margin} \cite{lepage2023ExperimentingAdditiveMargins,lepage2024AdditiveMargin} investigates the effect of margins, such as CosFace \cite{wang2018CosFace} and ArcFace \cite{deng2019ArcFace}, to improve the discriminative capacity of SSL contrastive losses by further separating positive from negative pairs in the representation space.

\paragraph{Channel-effects mitigation}
Through improved positive sampling strategies or additional regularization terms, several methods seek to disentangle speaker identity from channel factors related to recording conditions to improve the quality of speaker representations.
\textbf{AAT} \textit{(Augmentation Adversarial Training)} \cite{huh2020AAT} introduces an adversarial loss, penalizing the ability of the model to predict whether pairs of augmented segments share the same channel characteristics, to disentangle the speaker information from the channel information encoded in the representations. \textbf{SimCLR + AAT} \cite{tao2022LGL} employs the SimCLR framework and the \textbf{AAT} method \cite{huh2020AAT} to disentangle speaker and channel information. \textbf{DPP} \textit{(Diverse Positive Pairs)} \cite{tao2023DPP} introduces a multi-modal contrastive learning loss and aims to determine diverse positives from distinct utterances of the same speaker by cross-referencing speech and face data derived from training videos. \textbf{SimCLR-SSPS} \textit{(Self-Supervised Positive Sampling)} \cite{lepage2025BootstrappedPositiveSampling,lepage2025SSPS} addresses the limitation of SSL same-utterance positive sampling by finding appropriate pseudo-positives (i.e., same speaker identity but different recording condition) in latent space to reduce intra-speaker variance caused by extrinsic variabilities such as channel information.

\subsubsection{Self-distillation}

\paragraph{General methods}
Self-distillation, and particularly the DINO framework, became a central SSL approach in the field. It has been applied successfully across multiple encoder architectures, demonstrating strong performance and serving as a foundation for many subsequent studies.
\textbf{DINO (Fast ResNet-34)} \cite{cho2022DINO,cho2022DINO_2} presents a model following the DINO framework, which outperforms the MoCo contrastive baseline trained with a larger encoder. \textbf{DINO (RawNet3)} \cite{jung2022RawNet3} reports the performance of DINO used in combination with the RawNet3 encoder, an evolution of ECAPA-TDNN \cite{desplanques2020ECAPATDNN} and RawNet2 \cite{tak2021RawNet2}.

\paragraph{Regularization-based approaches}
Regularization strategies for self-distillation aim to stabilize training and improve embedding quality by leveraging auxiliary loss terms or initialization strategies, sometimes based on other SSL objectives or pre-trained SSL models. They generally enhance robustness and generalization but may require careful tuning of hyperparameters.
\textbf{SSReg} \textit{(Self-Supervised Regularization)} \cite{sang2022SSReg} paved the way for self-distillation approaches in SV by combining a contrastive learning objective function (AP loss \cite{chung2020DefenceMetricLearningSR}) and the SimSiam self-distillation framework as a regularization technique. \textbf{C3-DINO} \cite{zhang2022C3DINO} optimizes a DINO-based model, initialized from the weights of \textbf{C3-MoCo} \cite{zhang2022C3DINO}, described previously, to further improve downstream performance. \textbf{DINO + Cosine loss} \cite{han2022DLGLC} employs the DINO framework and a consistency regularization loss that maximizes the cosine similarity between embeddings from the same utterance, thereby encouraging consistent speaker representations in cosine space. \textbf{RDINO} \textit{(Regularized DINO)} \cite{chen2023RDINO} applies two regularization techniques to complement the DINO framework: a diversity term and a redundancy elimination term, based on the variance and covariance components of the VICReg objective function, respectively.

\paragraph{Augmentation-driven techniques}
Since the training data serves as both input and target in SSL, some methods enhance self-distillation frameworks by applying richer data-augmentations or diversifying input features to improve robustness to channel variability and other extrinsic factors.
\textbf{DINO + Aug.} \cite{chen2022ComprehensiveStudySelfDistillation} proposes a study on self-distillation approaches for SV and introduces a new extensive data-augmentation strategy based on pitch perturbation to further improve the robustness of DINO to channel perturbations. \textbf{RDINO + W-GVKT} \textit{(Within-Global-View Knowledge Transfer)} \cite{jin2024WGVKT} enables transferring knowledge from the teacher to the student based on the same global views and implements diversified versions of global views by applying spectral augmentation (SpecAugment \cite{park2019SpecAugment}) and feature diversification (MFCCs, LFCCs, spectrogram), to further facilitate the transfer of knowledge from the teacher to the student.

\paragraph{Curriculum learning}
Progressive training strategies gradually increase task difficulty or data diversity during self-distillation to stabilize training and improve convergence. They are particularly effective for managing large inter-speaker variability and reducing early-stage training instability.
\textbf{DINO + Curriculum} \cite{heo2022DINOCurriculum} shows that limiting inter-speaker variance at the initial phase of training is very effective and designs two curriculum learning strategies that: (1) gradually increases the number of speakers throughout the training; (2) gradually increases the proportion of augmented utterances.

\paragraph{Optimized positive sampling}
Through the selection of more informative positives, self-distillation can be improved by reducing intra-class variance introduced by extrinsic factors. These methods enhance the discriminative quality of representations and prove especially effective in conditions with high variability in recordings.
\textbf{CA-DINO} \textit{(Cluster-Aware DINO)} \cite{han2024CADINO} clusters speaker representations to sample positives from the same class as the anchor, providing an alternative to the conventional same-utterance positive sampling. \textbf{DINO-SSPS} \cite{lepage2025BootstrappedPositiveSampling,lepage2025SSPS} applies \textbf{SSPS} \textit{(Self-Supervised Positive Sampling)} to the DINO framework, similarly to \textbf{SimCLR-SSPS} \cite{lepage2025BootstrappedPositiveSampling}.

\paragraph{Prototype refinement}
Prototypes from the last linear layer of the projector are central to self-distillation, and several methods have investigated strategies to refine them or impose additional constraints to enhance representation compactness and inter-class separability.
\textbf{PDC-RDINO} \textit{(Prototype Division - Clustering)} \cite{zhao2024PrototypeDivisionDINO} generates fine-grained prototypes for the projection head by deriving new prototypes from the neighborhood of the existing prototypes to separate confused speaker classes. \textbf{SDPN} \textit{(Self-Distillation Prototypes Network)} \cite{chen2025SDPN} revisits the learning process of prototypes within the self-distillation framework, notably using Sinkhorn-Knopp \cite{cuturi2013Sinkhorn} and Me-Max regularization \cite{assran2021PAWS} to refine target distributions, and introduces a diversity regularization loss to maximize the inter-speaker distance of representations within the batch.

\paragraph{Distillation variants}
Several approaches extend the standard self-distillation framework by modifying the teacher-student interactions, for example through multiple branches, cross-distillation, or confidence-aware strategies.
\textbf{MeMo} \textit{(Multi-Head Multi-Mode)} \cite{jin2024MeMo} introduces an additional student-teacher pair to perform self-distillation between branches with the same architecture and cross-distillation between branches with different architectures, while applying contrastive learning on representations during the early training stages. \textbf{EBCA-DINO} \textit{(Energy-based Confidence-Aware Distillation)} \cite{hao2025EBCADINO} extends DINO by categorizing training samples into high- and low-energy groups based on their uncertainty, dynamically adjusting the teacher temperature to soften the target distribution for uncertain samples and sharpen the target distribution for confident samples.

\paragraph{Reconstruction-based learning}
Some methods have explored combining self-distillation with masked modeling objectives, where the model learns to infer missing frame-level representations from unmasked information, enhancing the contextual richness and robustness of the learned speaker embeddings.
\textbf{DINO + RMP} \textit{(Relational Mask Prediction)} \cite{kim2024RMP} proposes to combine DINO with a masked modeling objective on frame-level representations, through the introduction of a Block Aggregation Transformer (BA-Transformer) encoder, to capture global context information and enrich frame-level features.

\subsection{Multi-stage methods}

Multi-stage methods currently achieve state-of-the-art performance on SV by relying on the learned SSL speaker representations to provide supervision for a standard supervised classification loss. These approaches have the disadvantage of being more expensive to train, as they may require multiple iterations of pseudo-labels refinement, and the number of speaker classes $C$ must be estimated from the unlabeled data.

The training process is described as follows:
\begin{enumerate}
    \item Train a single-stage SSL method until convergence;
    \item Extract representations $\{\yi\}_{i \in \I}$ of all training samples;
    \item Cluster representations into $C$ classes and generate pseudo-labels $\{l_i\}_{i \in \I} \subset \{1, \dots, C\}$;
    \item Train a new model or fine-tune the encoder $\ftheta$ with a supervised classification loss using the pseudo-labels;
    \item Repeat steps 2, 3, and 4 for several iterations to refine the quality of the pseudo-labels (optional).
\end{enumerate}

For the supervised training, the joint-embedding architecture and the projector are discarded. A linear classifier is introduced to map representations to their respective speaker identities, such that $\boldsymbol{w}_c$ is the prototype of the $c$-th class. The standard objective function is the \textit{AAM-Softmax} loss \cite{deng2019ArcFace}, denoted by $\L[Sup.]$ and defined as:
\begin{equation}
    \label{eq:aamsoftmax_loss}
    \L[Sup.] = - \frac{1}{B} \sum_{i \in \B} \log \frac{\ell^{+}\left(\yi, \boldsymbol{w}_{l_i}\right)}{\ell^{+}\left(\yi, \boldsymbol{w}_{l_i}\right) + \sum\limits_{\substack{c = 1\\c \neq l_i}}^{C} \ell^{-}\left(\yi, \boldsymbol{w}_{c}\right)},
\end{equation}
where $\ell^{+}(\boldsymbol{a}, \boldsymbol{b})= \exp\left(s \cdot \cos\left(\theta_{\boldsymbol{a}, \boldsymbol{b}}+m\right)\right)$, $\ell^{-}(\boldsymbol{a}, \boldsymbol{b})= \exp\left(s \cdot \cos\left(\theta_{\boldsymbol{a}, \boldsymbol{b}}\right)\right)$, $\theta_{\boldsymbol{a},\boldsymbol{b}}$ is the angle between two feature vectors and $\cos(\theta_{\boldsymbol{a},\boldsymbol{b}})$ corresponds to their cosine similarity, $s$ is the scale hyperparameter, and $m$ is the margin hyperparameter.

Several methods have explored this training framework to further improve downstream performance. The most common approaches are presented below, classified by theme.

\paragraph{Methods without pseudo-label filtering}
Pseudo-labels can be used without any filtering or correction, relying directly on clustering outputs to iteratively refine the speaker representations. This straightforward approach simplifies the pipeline but may propagate clustering errors across iterations, especially in the presence of noisy data.
\textbf{IDLAB} \textit{(VoxSRC-20)} \cite{thienpondt2020IDLABVoxSRC20} uses the Sub-center AAM-Softmax loss \cite{deng2020SubcenterAAM}, to mitigate the effects of label noise and large intra-class variability, and increases the margin hyperparameter $m$ during the final training steps, which is referred to as \textbf{LMFT} \textit{(Large Margin Fine-Tuning)}.
\textbf{JHU} \textit{(VoxSRC-21)} \cite{cho2021JHUVoxSRC21} adopts the multi-stage training strategy starting from the single-stage \textbf{DINO (Fast ResNet-34)} \cite{cho2022DINO} method and applies \textbf{LMFT} without any particular filtering of pseudo-labels.
\textbf{SNU-HIL} \textit{(VoxSRC-21)} \cite{mun2021SNUHILVoxSRC21} follows the multi-stage training framework bootstrapped by the single-stage \textbf{CEL} \cite{mun2020CEL} method without any filtering of pseudo-labels.

\paragraph{Heuristic-based pseudo-label filtering}
Heuristic-based pseudo-label filtering methods aim to improve label reliability by removing low-confidence samples or small uncertain clusters. These straightforward yet effective strategies improve cluster purity and training stability, albeit at the cost of reduced data diversity due to the exclusion of potentially useful but ambiguous samples.
\textbf{DKU-DukeECE} \textit{(VoxSRC-20)} \cite{wang2020DKUDukeECEVoxSRC20} implements a strategy to purify the generated pseudo-labels by (1) discarding samples with the least clustering confidence (i.e., large distance between sample and centroid); (2) ignoring clusters for which the number of samples is insufficient.
\textbf{DukeECE} \cite{cai2021IterativeFramework} is an extension of the work in \textbf{DKU-DukeECE} \textit{(VoxSRC-20)} \cite{wang2020DKUDukeECEVoxSRC20} with a larger number of iterations.

\paragraph{Loss-based pseudo-label filtering}
Loss-based pseudo-label filtering represents one of the most common strategies to mitigate label noise, leveraging the training loss as an indicator of pseudo-label reliability. By adaptively or dynamically discarding high-loss samples, these methods improve robustness to noisy supervision, though they risk excluding challenging yet informative examples.
\textbf{LGL} \textit{(Loss Gated Learning)} \cite{tao2022LGL} is bootstrapped from the single-stage \textbf{SimCLR + AAT} \cite{tao2022LGL} method and discards unreliable pseudo-labels using a fixed threshold based on the finding that samples with incorrect pseudo-labels typically have higher loss values after a few training epochs, which can be attributed to the model's generalization to reliable pseudo-labels, representing the majority of the training data.
\textbf{DLG-LC} \cite{han2022DLGLC} is bootstrapped from the single-stage \textbf{DINO + Cosine loss} \cite{han2022DLGLC} method and extends \textbf{LGL} \cite{tao2022LGL} by filtering reliable and unreliable pseudo-labels: \textbf{DLG} \textit{(Dynamic Loss-Gate)} determines the filtering threshold by modeling the training loss distribution using a Gaussian Mixture Model (GMM) with two components (reliable and unreliable) and \textbf{LC} \textit{(Label Correction)} avoids discarding all unreliable pseudo-labels by minimizing the cross-entropy between the predicted posterior probability of clean and augmented segments of the same utterance when the predicted label shows high confidence.
\textbf{AT + HT} \cite{zhou2024ATHT} combines two components: \textbf{AT} \textit{(Adaptive Threshold)} to discard unreliable pseudo-labels similarly to \textbf{DLG} \cite{han2022DLGLC} and \textbf{HT} \textit{(Hierarchical Training)} to address the greater sensitivity of top layers to label noise by optimizing top layers after bottom layers have converged.

\paragraph{Pseudo-label correction}
Pseudo-label correction methods iteratively refine or adjust noisy pseudo-labels during training, combining filtering and model-based corrections to improve label reliability. These approaches can enhance downstream performance by retaining informative samples while reducing the negative impact of mislabeled data, though they may introduce additional computational complexity.
\textbf{Sub-PTM} \textit{(Sub-structure of Pre-Trained Model)} \cite{chen2023SubPTM} uses Infomap \cite{rosvall2008Infomap} for clustering and iteratively refines pseudo-labels by assigning more weight on the model prediction as the training progresses, rescaling reliable pseudo-labels in the loss, and discarding samples that do not have a consistent pseudo-label across chunks.
\textbf{SSRL} \textit{(Self-Supervised Reflective Learning)} \cite{cai2025SelfSupervisedReflective} enables continuous refinement of pseudo-labels during training with an online clustering mechanism, eliminating the need for additional clustering iterations, implements a label noise modeling strategy to give greater emphasis to reliable samples, and uses pseudo-label queues to filter out outlier predictions.
\textbf{BDS-BPLC} \textit{(Batch-scale Data Screening and Pseudo-Label Correction)} \cite{wang2025BDSBPLC} uses adaptive local and global thresholds to distinguish reliable from unreliable samples, applying a correction loss to the latter by leveraging the model's own predictions as supervision to refine pseudo-labels.
\textbf{AdaptiveDrop} \cite{fathan2025AdaptiveDrop} jointly integrates pseudo-label filtering and correction in an end-to-end framework by adaptively dropping or correcting samples based on their consistency with dynamically tracked dominant sub-centers per class, following Sub-center ArcFace \cite{deng2020SubcenterAAM} and BoundaryFace \cite{wu2022BoundaryFace}.

\paragraph{Cross-modal methods}
Cross-modal methods leverage complementary information from multiple modalities, typically audio and visual data, to generate more robust pseudo-labels through joint representations or knowledge transfer. These approaches can improve label quality and model generalization, especially in scenarios where one modality may be noisy or insufficient, but they require additional computational resources and careful handling of modality alignment.
\textbf{DKU-DukeECE} \textit{(VoxSRC-21)} \cite{cai2021DKUDukeECEVoxSRC21,cai2022IncorporatingVisualInformation} performs iterative clustering on multi-modal audio-visual representations, using a voting strategy to fuse the resulting assignments, and applies label smoothing regularization to deal with label noise.
\textbf{DPP} \cite{tao2023DPP} follows the standard multi-stage procedure starting from the single-stage \textbf{DPP} \cite{tao2023DPP} method and performs the clustering on concatenated speech-face representations.
\textbf{Co-Meta} \textit{(Co-teaching and Meta-learning)} \cite{chen2023SSLAudioVisualCoMeta} improves audio-visual multi-stage frameworks by transferring knowledge between modalities, following Co-teaching+ \cite{yu2019CoTeaching+} and MAML \cite{finn2017MAML}, such that the learning ability of the model on difficult samples can be improved by the other modality.
\textbf{CA-DINO + DLG-LC} \cite{han2024CADINO} applies the multi-stage training framework of \textbf{DLG-LC} \cite{han2022DLGLC} to \textbf{CA-DINO} \cite{han2024CADINO} and performs the clustering on the fusion of audio-visual representations.

\paragraph{Clustering algorithms}
Other approaches have introduced innovative clustering algorithms to produce robust pseudo-labels, enabling the training pipeline to operate in an end-to-end manner.
\textbf{CAMSAT} \textit{(Clustering based on Augmentation Mix and Self-Augmented Training)} \cite{fathan2024CAMSAT} learns robust pseudo-labels with a novel end-to-end SSL clustering algorithm based on IMSAT \cite{hu2017IMSAT}, which maximizes the mutual information between inputs and their discrete representation, and AugMix \cite{hendrycks2020AugMix}.

\paragraph{Encoders and base models}
Certain approaches leverage large-scale encoders to advance the performance frontier, while others generate initial pseudo-labels from lightweight unsupervised models to reduce computational complexity.
\textbf{DINO-WavLM} \cite{miara2024WavLMSSLSV} fine-tunes WavLM \cite{chen2022WavLM}, a large-scale SSL speech foundation model, with the MHFA back-end \cite{peng2022MHFA}, using pseudo-labels initially derived from a DINO-based model, and incorporates \textbf{DLG-LC} along with a final \textbf{LMFT} step.
\textbf{IPL} \textit{(Iterative Pseudo-Labeling)} \cite{aldeneh2025IPL} generates the initial pseudo-labels from the unsupervised i-vector method~\cite{dehak2011IVector} instead of a standard and more complex SSL model.
\section{Experimental setup}
\label{sec:exp_setup}

\subsection{Datasets and feature extraction}

Trainings are performed using the VoxCeleb2 \cite{chung2018VoxCeleb2} development set, which contains \num{1092009} utterances of \num{5994} speakers. During self-supervised training, speaker labels are discarded. Evaluation is performed on VoxCeleb1 \cite{nagrani2017VoxCeleb} \textit{original}, \textit{extended}, and \textit{hard} benchmarks, denoted by `VoxCeleb1-O', `VoxCeleb1-E', and `VoxCeleb1-H', respectively.

The extracted audio frames have a duration of \SI{2}{\second} by default. In the case of DINO, local frames correspond to four \SI{2}{\second} segments, while global frames consist of two \SI{4}{\second} segments. Audio segments are randomly extracted from an utterance with possible overlap. The input features are derived from $D_i$-dimensional log-mel spectrograms, computed using the torchaudio library with a Hamming window of \SI{25}{\milli\second} and a frame-shift of \SI{10}{\milli\second}. By default, the input dimension $D_i$ is set to 40 for all models, and to 80 for DINO-based models. Voice Activity Detection (VAD) is not employed, as training samples primarily contain continuous speech. Instance normalization is applied to normalize the input features.

\subsection{Data-augmentation}

SSL commonly uses data-augmentation to learn representations that are robust against extrinsic variabilities (e.g., environmental noise, mismatching recording devices, and varying acoustic conditions). Since anchor-positive pairs are derived from the same utterance, providing different augmented versions is fundamental to avoid encoding channel characteristics, ensuring that speaker identity is the main distinguishing factor.

Following the standard data-augmentation pipeline from the works on SSL for SV \cite{zhang2021SimCLR, xia2021MoCo}, two transformations are applied to the input waveforms at each training iteration: (1) reverberation and (2) background noise addition. First, reverberation is applied by randomly sampling an RIR from the Simulated Room Impulse Response Database \cite{ko2017StudyReverberantSpeechRobustSR}. Then, noises are added by randomly sampling background noises, music tracks, or speech segments from the MUSAN dataset \cite{snyder2015MUSAN}. To simulate a variety of acoustic scenarios, the Signal-to-Noise Ratio (SNR) is randomly sampled in the ranges \SIrange{0}{15}{\decibel} for noise, \SIrange{5}{15}{\decibel} for music, and \SIrange{13}{20}{\decibel} for speech.

For the DINO framework, which benefits from a more diverse data-augmentation strategy, the input is either left unaltered (no data-augmentation), augmented with reverberation or noise, or subjected to both transformations (similarly to the default strategy).

\subsection{SSL frameworks}

The encoder $\f$ is based on the Fast ResNet-34 \cite{chung2020DefenceMetricLearningSR} or the ECAPA-TDNN \cite{desplanques2020ECAPATDNN} architecture. The former (\SI{1.4}{\mega\text{ parameters}}) is used for hyperparameter tuning and the study of SSL components, while the latter (\SI{22.5}{\mega\text{ parameters}}) is used for the final evaluation. The Fast ResNet-34 model has a base hidden dimension of 16 and generates utterance-level representations using Self-Attentive Pooling (SAP) \cite{cai2018ExploringE2ESV}. The ECAPA-TDNN model has a hidden dimension of 1024 and relies on Attentive Statistics Pooling (ASP) \cite{okabe2018ASP}. Both encoders produce speaker representations of dimension $\Drepr=512$.

The projector $\g$ is an MLP consisting of three linear layers. Each layer, except the last one, is followed by a Batch Normalization and a ReLU activation function. The hidden dimension is set to 2048, and the final layer outputs embeddings of dimension $\Demb=512$. The projector is discarded for SimCLR and MoCo contrastive frameworks, since it negatively impacts downstream performance as detailed in Section~\ref{sec:study:projector}.

Models are optimized during 100 epochs using Adam, a batch size of 256, no weight decay, and a learning rate of 0.001, which is reduced by 5\% every 5 epochs. For DINO, the optimizer is SGD, the batch size is 128, the weight decay is $5e^{-5}$, and the learning rate is warmed up linearly to 0.2 during 10 epochs before using a decreasing cosine scheduler starting at 0.2. This specific training setup for DINO is discussed in more detail in Section~\ref{sec:hyperparams:dino:setup}.

The code is available as part of the \textit{sslsv}\footnote{\url{https://github.com/theolepage/sslsv}} toolkit based on the PyTorch framework. Trainings are conducted on 2 $\times$ NVIDIA Tesla V100 32 GB GPUs and 4 $\times$ NVIDIA Tesla V100 32 GB GPUs for DINO.

The implementation details for all SSL frameworks, along with the supervised baseline and the fusion system, are outlined below. The following hyperparameter settings are based on either the original implementations or prior adaptations to SV found in the literature. A systematic investigation into the main SSL hyperparameters of MoCo, SwAV, VICReg, and DINO is presented in the next section.

\paragraph{\textbf{Supervised}} The supervised baseline is trained using the AAM-Softmax loss of Eq.~(\ref{eq:aamsoftmax_loss}) with $s=30$ (scale) and $m=0.2$ (margin) in a supervised way with the training set labels.

\paragraph{\textbf{CPC}} The number of timesteps $T$ is 51 (Fast ResNet-34 without pooling) and $t_0$ is set to $T-4$. The aggregator $g_{\phi}$ is a GRU with dimension 256. The predictor $h_{\psi}$  consists of a linear layer for each timestep to be predicted. The loss temperature is set to $\tau=1.0$.
\paragraph{\textbf{SimCLR}} The loss used is the symmetric NT-Xent, with a temperature of $\tau=0.03$.
\paragraph{\textbf{MoCo}} The loss temperature is $\tau=0.03$ and the memory queue has a length of $|\boldsymbol{Q}^{\text{MoCo}}|=\num{32768}$. The coefficient $m$ for the EMA update is fixed at $m=0.999$.

\paragraph{\textbf{DeepCluster}} The loss temperature is $\tau=0.1$. The number of clusters is set to $K=\num{3000}$.
\paragraph{\textbf{SwAV}} The loss temperature is $\tau=0.1$. The number of prototypes is $P=\num{6000}$. The queue of length $|\boldsymbol{Q}^{\text{SwAV}}|=\num{3840}$ is enabled after 15 epochs. The prototypes remain frozen during the first epoch.

\paragraph{\textbf{W-MSE}} The output dimension of the projector is set to $\Demb = 64$. The whitening is applied once and the window size for batch slicing is 128.
\paragraph{\textbf{Barlow Twins}} The scaling factor $\lambda$ for the redundancy reduction term is set to $0.005$.
\paragraph{\textbf{VICReg}} The loss scaling hyperparameters are set to $\lambda=1$, $\mu=1$, and $\nu=0.1$ for the invariance, variance, and covariance terms, respectively.

\paragraph{\textbf{BYOL}} The predictor $h_\psi$ consists of a linear layer, followed by Batch Normalization and ReLU, and a final linear layer. The hidden dimension of the predictor is set to 4096. The coefficient $m$ of the EMA update increases from 0.996 to 1.0 using a cosine scheduler.
\paragraph{\textbf{SimSiam}} A Batch Normalization module is added at the end of the projector. The predictor $h_\psi$ is similar to BYOL except that the hidden dimension is set to $512$.
\paragraph{\textbf{DINO}} By default, the number of global and local views is $G=2$ and $L=4$. The dimension of the last layer of the projector is set to 256 to create a bottleneck, and a final layer is added to map $l_2$-normalized embeddings to $\Demb=\num{65536}$ units. This final layer uses weight normalization and is not optimized during the first epoch. The student and teacher temperatures are fixed to $\tau_{\text{s}}=0.1$ and $\tau_{\text{t}}=0.04$, respectively. The coefficient $m$ of the EMA update increases from 0.996 to 1.0 following a cosine scheduler. Gradients with a norm exceeding 3.0 are clipped.

\paragraph{\textbf{Fusion}} The score-level fusion is trained using a logistic regression model optimized with LBFGS for 50 epochs on a custom trials list generated on VoxCeleb2. It integrates SimCLR, MoCo, SwAV, VICReg, and DINO models based on the ECAPA-TDNN encoder.

\subsection{Evaluation protocol}

To evaluate the performance on SV, the final model weights are obtained by averaging those from the last 10 epochs, ensuring more consistent results. Representations are extracted by processing full-length test audio samples. The trial score is determined by computing the cosine similarity of the pair of $l_2$-normalized speaker representations.

The performance is measured by the Equal Error Rate (EER) and minimum Detection Cost Function (minDCF), following the NIST Speaker Recognition Evaluation (SRE) protocol \cite{sadjadi2017NISTSRE2016}. Specifically, this article reports \mindcf, computed using $P_{target} = 0.01$, $C_{\text{miss}}=1$ and $C_{\text{fa}}=1$. For comparison purposes, $\text{minDCF}_\text{0.05}$ is also reported when methods from the literature provide results with $P_{target} = 0.05$ instead of $P_{target} = 0.01$.

\section{Effect of SSL hyperparameters}
\label{sec:hyperparams}

\subsection{Contrastive learning (MoCo)}
\label{sec:hyperparams:moco}

\begin{table}[t]
  \footnotesize
  \titlecaption{MoCo hyper-parameters --- Effect of the number of negatives ($|\boldsymbol{Q}^{\textnormal{MoCo}}|$) and class-collisions on \ac{SV} performance}{The encoder is Fast ResNet-34 and the benchmark is VoxCeleb1-O. Class-collisions are avoided using labels from the training set. The selected configuration is \colorbox{SelectedRowBg}{highlighted}.}
  \label{tab:moco_negs}
  \centering
  \color{SecondaryRowFg}
  \arrayrulecolor{PrimaryRowFg} 
  \begin{tabular}{S[table-format=5.0]S[table-format=1.2]S[table-format=1.4]S[table-format=1.2]S[table-format=1.4]}
    \toprule
    \color{PrimaryRowFg} \multirow{2}{*}{\shortstack[c]{\\\textbf{Num. negs.}\\\text{ ($|\boldsymbol{Q}^{\text{MoCo}}|$)}}} & \multicolumn{2}{c}{\color{PrimaryRowFg}\textit{Default (SSL)}} & \multicolumn{2}{c}{\color{PrimaryRowFg}\textit{No false-negs. (Sup.)}} \\
    \cmidrule(lr){2-3} \cmidrule(lr){4-5}
    & \color{PrimaryRowFg} \textbf{\eer} & \color{PrimaryRowFg} \textbf{\mindcf} & \color{PrimaryRowFg} \textbf{\eer} & \color{PrimaryRowFg} \textbf{\mindcf} \\
    \midrule
    256 & 9.35 & 0.6513 & 9.17 & 0.6426 \\
    1024 & 9.30 & 0.6604 & 8.69 & 0.6111 \\
    16384 & 8.92 & 0.6235 & 8.33 & \bfseries 0.5790 \\
    \rowcolor{SelectedRowBg} \color{PrimaryRowFg} 32768 & \color{PrimaryRowFg} \bfseries 8.49 & \color{PrimaryRowFg} \bfseries 0.5990 & \bfseries 7.92 & 0.5904 \\
    65536 & 8.71 & 0.6044 & 8.50 & 0.6197 \\
    \bottomrule
  \end{tabular}
\end{table}

\begin{table}[t]
  \footnotesize
  \titlecaption{MoCo hyper-parameters --- Effect of the temperature ($\tau$) on \ac{SV} performance}{The encoder is Fast ResNet-34 and the benchmark is VoxCeleb1-O. The selected configuration is \colorbox{SelectedRowBg}{highlighted}.}
  \label{tab:moco_temp}
  \centering
  \color{SecondaryRowFg}
  \arrayrulecolor{PrimaryRowFg} 
  \begin{tabular}{S[table-format=1.2]S[table-format=1.2]S[table-format=1.4]S[table-format=1.2]}
    \toprule
    \color{PrimaryRowFg} {\textbf{Temperature} ($\tau$)} & \color{PrimaryRowFg} \textbf{\eer} & \color{PrimaryRowFg} \textbf{\mindcf} & \color{PrimaryRowFg} \textbf{Entropy} \\
    \midrule
    0.01 & \bfseries 8.25 & 0.6218 & 0.9 \\
    \rowcolor{SelectedRowBg} \color{PrimaryRowFg} 0.03 & \color{PrimaryRowFg} 8.49 & \color{PrimaryRowFg} \bfseries 0.5990 & 2.2 \\
    0.05 & 8.44 & 0.6124 & 4.1 \\
    0.07 & 8.80 & 0.6356 & 6.4 \\
    0.1 & 9.63 & 0.6935 & 8.4 \\
    \bottomrule
  \end{tabular}
\end{table}

\begin{table}[t]
  \footnotesize
  \titlecaption{MoCo hyper-parameters --- Effect of the momentum coefficient ($m$) on \ac{SV} performance}{The encoder is Fast ResNet-34 and the benchmark is VoxCeleb1-O. The selected configuration is \colorbox{SelectedRowBg}{highlighted}.}
  \label{tab:moco_momentum}
  \centering
  \color{SecondaryRowFg}
  \arrayrulecolor{PrimaryRowFg} 
  \begin{tabular}{S[table-format=1.3]S[table-format=2.2]S[table-format=1.4]}
    \toprule
    \color{PrimaryRowFg} {\textbf{Momentum} ($m$)} & \color{PrimaryRowFg} \textbf{\eer} & \color{PrimaryRowFg} \textbf{\mindcf} \\
    \midrule
    0.9 & 24.36  & 0.9997 \\
    0.99 & 8.37 & 0.6050 \\
    0.996 & \bfseries 8.20 & 0.6169 \\
    \rowcolor{SelectedRowBg} \color{PrimaryRowFg} 0.999 & \color{PrimaryRowFg} 8.49 & \color{PrimaryRowFg} \bfseries 0.5990 \\
    1.0 & 24.47 & 0.9839 \\
    \bottomrule
  \end{tabular}
\end{table}

For contrastive learning frameworks, the effect of MoCo hyperparameters on SV performance is assessed, notably the number of negatives, the loss temperature, and the momentum update coefficient.

\subsubsection{Number of negatives}
Table~\ref{tab:moco_negs} shows the effect of the number of negatives, which is determined by the size of the queue used by MoCo (i.e., the $|\boldsymbol{Q}^{\text{MoCo}}|$ hyperparameter), on SV performance. The number of negatives has a negligible impact on downstream performance, as the model achieves 9.35\% and 8.71\% EER with the minimum (256) and maximum (\num{65536}) number of negatives, respectively. Note that the former configuration is comparable to the SimCLR framework, which uses 255 negatives since the training batch size is set to 256. Moreover, the performance saturates with a larger number of negatives. The optimal configuration, highlighted in blue, is obtained with $|\boldsymbol{Q}^{\text{MoCo}}|=\num{32768}$, yielding 8.49\% EER and 0.5990 minDCF.

\subsubsection{Class collisions}
Table~\ref{tab:moco_negs} also illustrates the impact of class collisions by reporting the results obtained when discarding false-negative contrastive pairs using the training set labels. Class collisions have a relatively small effect on SV performance since canceling false-negative pairs only leads to an EER of 7.92\% with the best configuration, which is a relative improvement of 6.7\%.

The probability of a class collision (i.e., sampling at least one false negative), assuming that classes are balanced and that negatives are sampled uniformly at random, can be expressed as $P_{\text{collision}} = 1 - \left(1 - \frac{1}{C}\right)^{n}$, where $n$ is the number of negatives and $C$ is the number of classes. This probability increases as the number of negatives $n$ grows, and decreases as the number of classes $C$ grows, because a higher class count reduces the chance of randomly sampling the anchor’s class. With the present experimental setup ($C=5994$ and $n=\num{32768}$), the probability of having at least one class collision among the negatives is approximately 99.6\%, but false-negative samples only account for approximately 0.017\% of all anchor-negative pairs.

These findings suggest that having a very large number of negatives is not necessary and that class collisions have a negligible impact on training convergence, although several works have been proposed to address this issue \cite{xia2021MoCo,zhang2022C3DINO}.

\subsubsection{Temperature}
Table~\ref{tab:moco_temp} shows the role of the temperature $\tau$ used in the loss of Eq.~(\ref{eq:moco_loss}). Tuning this hyperparameter is critical to achieving the best downstream performance, as it controls the smoothness of the similarity distribution, which is highlighted by the entropy values. In particular, a lower $\tau$ sharpens the distribution, making the model focus more on hard negatives (i.e., negatives that are very similar to the anchor). The best result in terms of minDCF, highlighted in blue, achieves 8.49\% EER and 0.5990 minDCF with $\tau=0.03$. Note that this is approximately equivalent to the scale $s=30$ commonly used in the supervised setting with the AAM-Softmax loss, as defined in Eq.~(\ref{eq:aamsoftmax_loss}).

\subsubsection{Momentum update}
Table~\ref{tab:moco_momentum} highlights the effect of the momentum update coefficient $m$. The momentum update is necessary to prevent collapse, as it ensures that the key branch is updated gradually with respect to the query branch, while the coefficient $m$ controls the rate of this update. Notably, a higher value of $m$ ensures that $\theta^{\prime}$ updates slowly, providing stable negative samples for contrastive learning by avoiding the moving target problem (i.e., where the negative samples are constantly changing). The optimal configuration in terms of minDCF, highlighted in blue, is obtained with $m=0.999$ and yields 8.49\% EER and 0.5990 minDCF.

\subsection{Clustering (SwAV)}
\label{sec:hyperparams:swav}

\begin{table}[t]
  \footnotesize
  \titlecaption{SwAV hyper-parameters --- Effect of the number of prototypes ($P$) on \ac{SV} performance}{The encoder is Fast ResNet-34 and the benchmark is VoxCeleb1-O. The selected configuration is \colorbox{SelectedRowBg}{highlighted}.}
  \label{tab:swav_nbproto}
  \centering
  \color{SecondaryRowFg}
  \arrayrulecolor{PrimaryRowFg}
  \begin{tabular}{
    S[table-format=5.0]
    S[table-format=2.2]
    S[table-format=1.4]
  }
    \toprule
    \color{PrimaryRowFg} \textbf{Num. prototypes} \text{ ($P$)} & \color{PrimaryRowFg} \textbf{\eer} & \color{PrimaryRowFg} \textbf{\mindcf} \\
    \midrule
    1000 & 12.12 & 0.7200 \\
    3000 & 12.28 & 0.7292 \\
    \rowcolor{SelectedRowBg} \color{PrimaryRowFg} 6000 & \color{PrimaryRowFg} \bfseries 11.82 & \color{PrimaryRowFg} \bfseries 0.7177 \\
    9000 & 12.31 & 0.7291 \\
    12000 & 13.30 & 0.7379 \\
    \bottomrule
  \end{tabular}
\end{table}

Regarding frameworks based on clustering, the role of the main hyperparameter of SwAV on SV performance is evaluated, as shown in Table~\ref{tab:swav_nbproto}. Specifically, the hyperparameter considered is the number of prototypes $P$ used by the online clustering algorithm during the training. The best configuration, highlighted in blue, reaches 11.82\% EER and 0.7177 minDCF when setting $P=6000$. This value is consistent with the number of actual speaker identities in the training set, as VoxCeleb2 comprises \num{5994} speakers.

\subsection{Information maximization (VICReg)}
\label{sec:hyperparams:vicreg}

\begin{table}[t]
  \footnotesize
  \titlecaption{VICReg hyper-parameters --- Effect of the loss weights ($\lambda$, $\mu$, and $\nu$) on \ac{SV} performance}{The encoder is Fast ResNet-34 and the benchmark is VoxCeleb1-O. The selected configuration is \colorbox{SelectedRowBg}{highlighted}.}
  \label{tab:vicreg_weights}
  \centering
\color{SecondaryRowFg}
  \arrayrulecolor{PrimaryRowFg}
  \begin{tabular}{
    S[table-format=1.0]
    S[table-format=1.1]
    S[table-format=1.2]
    S[table-format=2.2]
    S[table-format=1.4]
  }
    \toprule
    \color{PrimaryRowFg} \textbf{Inv.} \text{ ($\lambda$)} & \color{PrimaryRowFg} \textbf{Var.} \text{ ($\mu$)} & \color{PrimaryRowFg} \textbf{Cov.} \text{ ($\nu$)} & \color{PrimaryRowFg} \textbf{\eer} & \color{PrimaryRowFg} \textbf{\mindcf} \\
    \midrule
    1 & 1 & 0 & 27.24 & 0.9941 \\
    1 & 0.5 & 0.1 & 12.75 & 0.6893 \\
    1 & 1 & 0.04 & 11.94 & 0.6886 \\
    \rowcolor{SelectedRowBg} \color{PrimaryRowFg} 1 & \color{PrimaryRowFg} 1 & \color{PrimaryRowFg} 0.1 & \color{PrimaryRowFg} \bfseries 11.33 & \color{PrimaryRowFg} \bfseries 0.6658 \\
    \bottomrule
  \end{tabular}
\end{table}

For information maximization frameworks, the effect of VICReg hyperparameters on SV performance is investigated, as shown in Table~\ref{tab:vicreg_weights}. The hyperparameters considered are the invariance ($\lambda$), variance ($\mu$), and covariance ($\nu$) weights of the objective function defined in Eq.~(\ref{eq:vicreg_loss}). The weight of the invariance term is set to $\lambda=1$ for all configurations, as it is essential for learning representations robust to the different transformations simulated with data-augmentation. Performance degrades when reducing the weight of the variance term (second and fourth rows), which highlights the importance of this component in generating class-dependent representations and preventing a collapse. The covariance term also has a significant impact on the results, as it leads to a 58.4\% EER improvement between the first and last configurations. The optimal configuration, highlighted in blue, achieves the best performance with $\lambda=1$, $\mu=1$, and $\nu=0.1$, which yields an EER of 11.33\% and a minDCF of 0.6658. Notably, this configuration differs from the hyperparameters originally proposed for CV applications \cite{bardes2022VICReg}, suggesting that domain-specific tuning is necessary.

\subsection{Self-distillation (DINO)}
\label{sec:hyperparams:dino}

For self-distillation frameworks, the effect of DINO training setup and hyperparameters on SV performance is assessed, notably the frame sampling, projector output dimension, and teacher temperature.

\subsubsection{Training setup}
\label{sec:hyperparams:dino:setup}
\newcommand{\printN}[1]{
    \ifnum#1>0
        \hspace{1pt}%
        \printN{\numexpr#1-1\relax}%
    \fi
}

\begin{table}[t]
  \footnotesize
  \titlecaption{DINO hyper-parameters --- Effect of the training setup on \ac{SV} performance}{The encoder is Fast ResNet-34 and the benchmark is VoxCeleb1-O. The selected configuration is \colorbox{SelectedRowBg}{highlighted}.}
  \label{tab:dino_setup}
  \centering
  \begin{tabular}{
    l
    S[table-format=2.2]
    S[table-format=1.4]
  }
    \toprule
    \textbf{Method} & \textbf{\eer} & \textbf{\mindcf} \\
    \midrule
    DINO & 19.24 & 0.9781 \\
    \printN{1} + Weight decay \scriptsize{($\text{wd}=5e^{-5}$)} & 20.01 & 0.9456 \\
    \printN{2} + LR sched. \scriptsize{(warmup + cosine)} & 10.47 & 0.6620 \\
    \printN{3} + Encoder pooling \scriptsize{(ASP)} & 8.85 & 0.6160 \\
    \printN{4} + Data-aug. \scriptsize{(clean/rir/noise/both)} & 7.00 & 0.5449 \\
    \printN{5} + Input dim. \scriptsize{($D_i=80$)} & \bfseries 5.94 & 0.4903 \\
    \rowcolor{SelectedRowBg} \printN{6} + Optimizer \scriptsize{(SGD, $\text{lr}=0.2$)} & 6.04 & \bfseries 0.4526 \\
    \bottomrule
  \end{tabular}
\end{table}
In the different implementations from the literature, DINO relies on a distinct training setup compared to the other frameworks. Table~\ref{tab:dino_setup} presents the incremental effect of each modification, starting from the initial setup (first row) to the optimal configuration in terms of minDCF (last row, highlighted in blue). The first row corresponds to the default experimental setup described for all SSL frameworks, while the last row corresponds to the specific setup used for the DINO framework, both detailed in Section~\ref{sec:exp_setup}. Specifically, DINO benefits from a weight decay and learning rate scheduler, a more diversified data-augmentation strategy, an increased input dimensionality, and the use of SGD as the optimizer. Using Attentive Statistics Pooling (ASP) \cite{okabe2018ASP} instead of Self-Attentive Pooling (SAP) \cite{cai2018ExploringE2ESV} also improves the convergence, which only applies to the Fast ResNet-34 encoder, since ASP is the default pooling for the ECAPA-TDNN encoder. These modifications do not yield the same improvements for other frameworks, suggesting that DINO is particularly sensitive to training conditions and that such adjustments are necessary to fully exploit its capacity for learning from bootstrapped supervisory signals. Similar adaptations of this experimental setup are also found in DINO-based methods from the literature, indicating a broader consensus on the importance of these design choices to achieve competitive performance \cite{cho2022DINO,chen2023RDINO}.

\subsubsection{Frame sampling}
\begin{table}[t]
  \footnotesize
  \titlecaption{DINO hyper-parameters --- Effect of frame sampling ($G$, $L$ and their respective durations) on \ac{SV} performance}{The encoder is Fast ResNet-34 and the benchmark is VoxCeleb1-O. The selected configuration is \colorbox{SelectedRowBg}{highlighted}.}
  \label{tab:dino_frames}
  \centering
  \begin{tabular}{
    c
    c
    S[table-format=1.2]
    S[table-format=1.4]
  }
    \toprule
    \textbf{Global views} ($G$) & \textbf{Local views} ($L$) & \textbf{\eer} & \textbf{\mindcf} \\
    \midrule
    1 $\times$ \SI{2}{\second} & 1 $\times$ \SI{2}{\second} & 14.53 & 0.7636 \\
    \midrule
    \rowcolor{SelectedRowBg} 2 $\times$ \SI{4}{\second} & 4 $\times$ \SI{2}{\second} & \bfseries 6.04 & \bfseries 0.4526 \\
    2 $\times$ \SI{3}{\second} & 4 $\times$ \SI{2}{\second} & 6.61 & 0.5195 \\
    2 $\times$ \SI{2}{\second} & 4 $\times$ \SI{2}{\second} & 8.33 & 0.6543 \\
    1 $\times$ \SI{2}{\second} & 4 $\times$ \SI{2}{\second} & 9.31 & 0.6793 \\
    \midrule
    2 $\times$ \SI{4}{\second} & 2 $\times$ \SI{2}{\second} & 6.43 & 0.4855 \\
    \bottomrule
  \end{tabular}
\end{table}

Table~\ref{tab:dino_frames} shows the impact of the frame sampling strategy, specifically the number of global and local views, $G$ and $L$, and their respective duration. The first row corresponds to the default setup used in all other frameworks, consisting of 2 segments of \SI{2}{\second} each. This configuration results in degraded performance, indicating the importance of the multi-crop strategy as a core component of DINO. Subsequent rows illustrate the sensitivity of the framework to both the number and duration of global views: reducing the length to \SI{3}{\second} or \SI{2}{\second}, or limiting the number to a single global view, consistently leads to performance degradation. Additionally, the last row shows that decreasing the number of local views from 4 to 2 also negatively impacts performance. The optimal configuration, highlighted in blue, involves two global views of \SI{4}{\second} and four local views of \SI{2}{\second}, achieving the best results with an EER of 6.04\% and a minDCF of 0.4526.

\subsubsection{Projector output dimension}
\begin{table}[t]
  \footnotesize
  \titlecaption{DINO hyper-parameters --- Effect of the projector output dimension ($D_{\textnormal{e}}$) on \ac{SV} performance}{The encoder is Fast ResNet-34 and the benchmark is VoxCeleb1-O. The selected configuration is \colorbox{SelectedRowBg}{highlighted}.}
  \label{tab:dino_K}
  \centering
  \begin{tabular}{
    S[table-format=5.0]
    S[table-format=1.2]
    S[table-format=1.4]
  }
    \toprule
    \textbf{Head output dim.} \text{ ($\Demb$)} & \textbf{\eer} & \textbf{\mindcf} \\
    \midrule
    2048 & 7.59 & 0.5189 \\
    16384 & 6.53 & 0.4848 \\
    32768 & 6.71 & 0.4921 \\
    \rowcolor{SelectedRowBg} 65536 & \bfseries 6.04 & \bfseries 0.4526 \\
    \bottomrule
  \end{tabular}
\end{table}

Table~\ref{tab:dino_K} investigates the effect of the projector output dimensionality $\Demb$. Increasing the dimension of embeddings from 2048 to \num{16384} significantly improves both the EER and the minDCF, suggesting that higher-dimensional projections are beneficial. However, further increasing the dimension to \num{32768} does not lead to continued improvements, highlighting that the effect of this hyperparameter is not linear and that increasing the projector output dimension indiscriminately does not guarantee better results. The best configuration is obtained with an output dimension of \num{65536}, highlighted in blue, yielding an EER of 6.04\% and a minDCF of 0.4526. This experiment shows that the projector output dimension plays a crucial role in the training process at the heart of self-distillation, as a larger dimensional space facilitates the emergence of more stable and diverse target distributions.

\subsubsection{Teacher temperature}
\begin{table}[t]
  \footnotesize
  \titlecaption{DINO hyper-parameters --- Effect of the teacher temperature ($\tau_{\textnormal{t}}$) on \ac{SV} performance}{The encoder is Fast ResNet-34 and the benchmark is VoxCeleb1-O. The selected configuration is \colorbox{SelectedRowBg}{highlighted}.}
  \label{tab:dino_temp}
  \centering
  \begin{tabular}{cS[table-format=1.2]S[table-format=1.4]S[table-format=1.2]}
    \toprule
    {\textbf{Teacher temp.} ($\tau_{\text{t}}$)} & \textbf{\eer} & \textbf{\mindcf} & \textbf{Entropy} \\
    \midrule
    0.01 & 23.18 & 0.9989 & 0.02 \\
    0.03 & 7.42 & 0.4948 & 0.04 \\
    \rowcolor{SelectedRowBg} 0.04 & 6.04 & \bfseries 0.4526 & 0.07 \\
    0.05 & 6.38 & 0.4753 & 0.28 \\
    0.07 & \bfseries 5.61 & 0.4567 & 3.18 \\
    \midrule
    0.04 $\rightarrow$ 0.07 & 6.25 & 0.4858 & 3.90 \\
    \bottomrule
  \end{tabular}
\end{table}

Table~\ref{tab:dino_temp} presents the role of the teacher temperature $\tau_{\text{t}}$ used in the loss of Eq.~(\ref{eq:dino_loss}). Similarly to MoCo, tuning this hyperparameter is critical to achieving the best downstream performance, as it controls the smoothness of the target distribution, as shown by the entropy values. Specifically, a low $\tau_{\text{t}}$ produces sharper, more confident predictions that can lead to training instability or collapse, while a high $\tau_{\text{t}}$ results in softer targets that may hinder the learning of discriminative representations. Using $\tau_{\text{t}}=0.01$ leads to 23.18\% EER, indicating a collapsing solution. Note that the bottom part of the table shows a configuration where the teacher temperature is linearly increased during the first 10 training epochs, which corresponds to the setup used in the original implementation of DINO for CV \cite{caron2021DINO}. This warming-up strategy aims to help stabilize the early training phase by avoiding overly confident targets from the teacher while progressively encouraging sharper distributions for better alignment with the student. The best result in terms of minDCF, highlighted in blue, achieves 6.04\% EER and 0.4526 minDCF with $\tau=0.04$ fixed throughout the training.
\section{Study of SSL components}
\label{sec:study}

To better understand the behavior and limitations of SSL frameworks on SV, the key components that influence model performance are analyzed. This includes investigating the risk of collapse, the role of data-augmentation and training data distribution, the function of the projector, and the impact of positive sampling. Each component is studied across multiple SSL frameworks to highlight common patterns, identify dependencies, and inform practical design choices for future applications.

\subsection{Collapse}
\label{sec:study:collapse}
\begin{figure}[t]
    \centering
    \includegraphics[height=0.3cm]{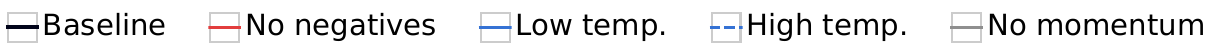}\\[0.15em]
    \subfloat[\textbf{MoCo}\label{fig:collapse:moco}]{
        \centering
        \includegraphics[width=0.49\linewidth]{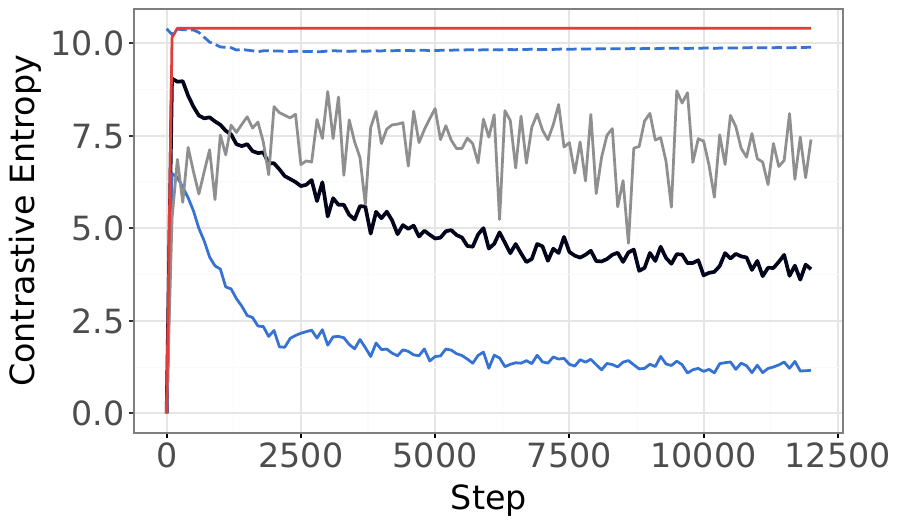}
        \includegraphics[width=0.49\linewidth]{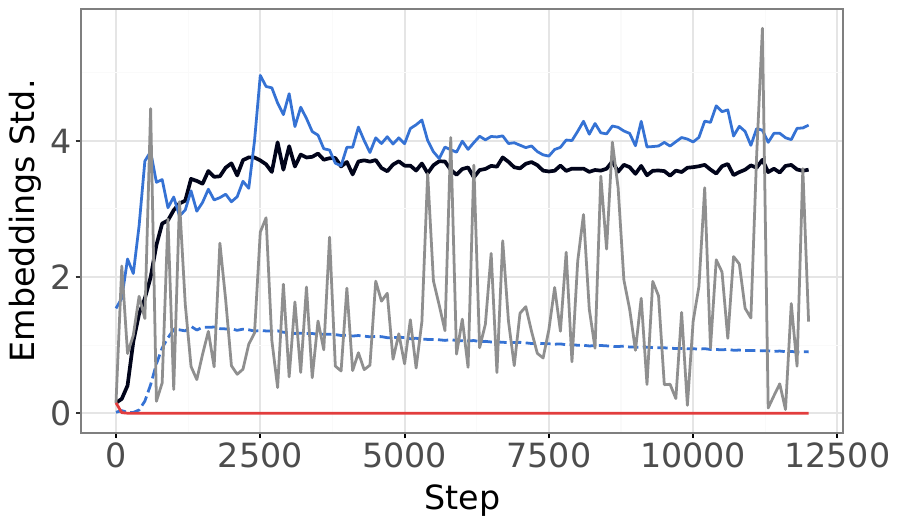}
    }\\[1em]
    \includegraphics[height=0.33cm]{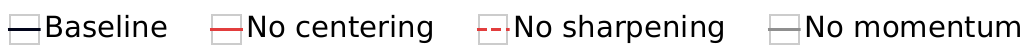}\\[0.15em]
    \subfloat[\textbf{DINO}\label{fig:collapse:dino}]{
        \centering
        \includegraphics[width=0.49\linewidth]{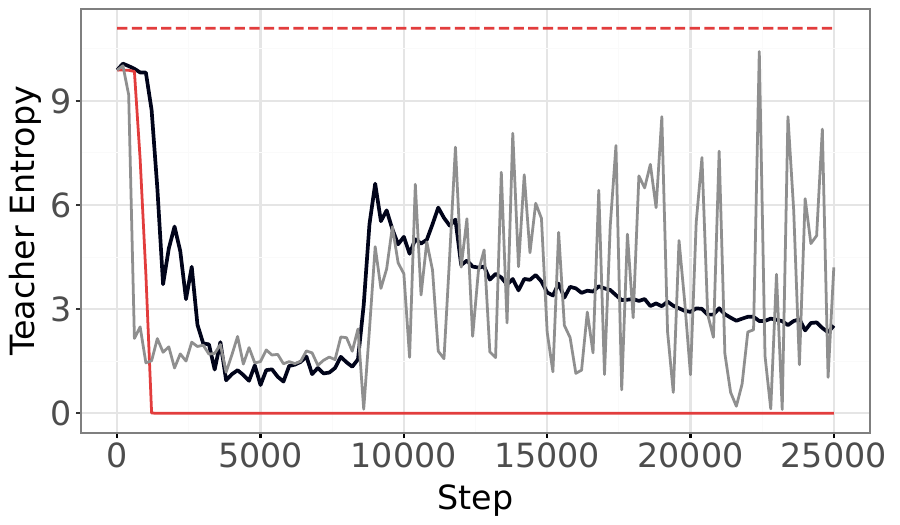}
        \includegraphics[width=0.49\linewidth]{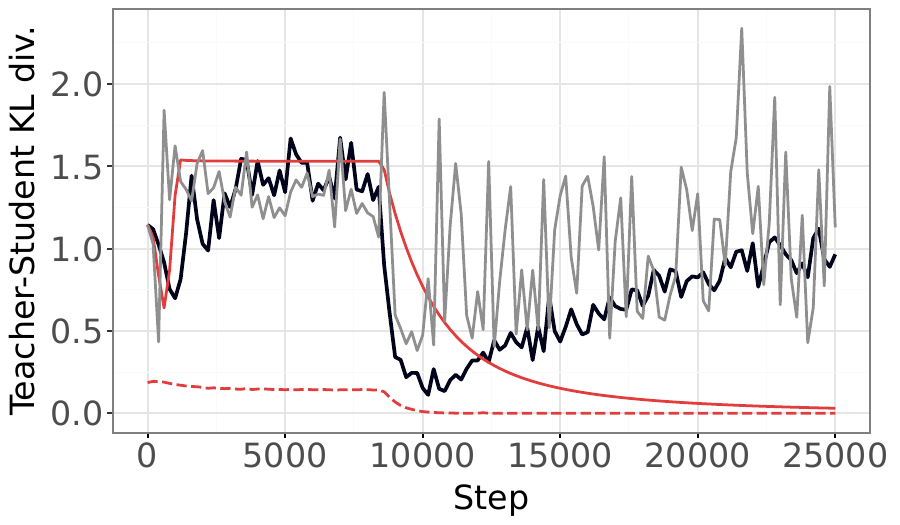}
    }
    \titlecaption{Collapse study of MoCo (a) and DINO (b) \ac{SSL} frameworks}{For MoCo (a), the entropy of the contrastive distribution and the standard deviation of embeddings are reported. For DINO (b), the entropy of the teacher distribution and the KL divergence of the teacher and student distributions are reported. Metrics are reported throughout training iterations of the first 3 epochs. The encoder is Fast ResNet-34.}
    \label{fig:collapse}
\end{figure}

Collapse refers to a degeneration of the learned representations, where embeddings become either indistinguishable across samples or restricted to a low-dimensional subspace. Most SSL frameworks rely on maximizing the similarity between two views of the same input data, which can lead to collapse if not adequately controlled. Therefore, each SSL paradigm employs specific mechanisms to prevent this issue.

In contrastive frameworks such as MoCo, collapse is prevented by the use of negative pairs, represented in the denominator of Eq.~(\ref{eq:moco_loss}). Figure~\subref*{fig:collapse:moco} illustrates the evolution of the entropy of the contrastive distribution and the standard deviation of embeddings using different configurations. The effect of the negative sampling mechanism is reflected in the entropy of the contrastive distribution, which quickly converges to $-log(1/|\boldsymbol{Q}^{\textnormal{MoCo}}|) \approx 10.40$ during the first iterations, when no negative pairs are created. This indicates a nearly uniform softmax distribution across all samples because the model produces constant embeddings. The temperature parameter significantly impacts the sharpness of this distribution: a low temperature yields more confident (lower entropy) predictions, while a high temperature results in less confident (higher entropy) predictions. The standard deviation of embeddings further reveals that collapse-prone configurations, such as ``No negatives'' and ``High temp.'', produce near-constant embeddings, as indicated by the standard deviation tending toward zero. Additionally, disabling momentum updates of the weights from the key branch leads to significant instability and collapse as the key encoder becomes unstable. This leads the model to exploit inconsistencies (shortcut cues) between recent positives and stale negatives, causing the stored negatives to drift and ultimately undermining the learning process.

In self-distillation frameworks such as DINO, collapse is notably prevented with normalization techniques, specifically centering and sharpening of the output distributions. Figure~\subref*{fig:collapse:dino} shows the evolution of the entropy of the teacher distribution and the KL divergence of the teacher and student distributions, for different configurations. Two distinct forms of collapse can occur: informational collapse, where the model output is uniform across all embedding dimensions, and dimensional collapse, where the output is dominated by a single dimension or a low-dimensional subspace. When centering is disabled, the entropy converges to zero, indicating a nearly one-hot distribution consistent with dimensional collapse. In contrast, when sharpening is disabled, the entropy approaches the maximum value of $-log(1/\Demb) \approx 11.09$, indicating an almost uniform distribution consistent with informational collapse. The Kullback-Leibler (KL) divergence between the teacher and student output distributions:
\begin{equation}
    D_{\mathrm{KL}}(P_{\text{t}}\,\|\,P_{\text{s}})=\sum_{x \in \mathcal{X}} P_{\text{t}}(x) \log \frac{P_{\text{t}}(x)}{P_{\text{s}}(x)},
\end{equation}
where $P_{\text{t}}$ is the teacher probability distribution and $P_{\text{s}}$ is the student probability distribution, confirms that both ablations lead to a constant output, hence a collapse. Moreover, removing the asymmetry between the student and teacher branches results in strong instability and collapse, since momentum updates are essential to maintain a stable and slowly evolving teacher that provides consistent targets leading to stable training signals. Note that the sudden change in values at the 8531-th training step corresponds to the unfreezing of the last layer of the projector.

\subsection{Data-augmentation}
\label{sec:study:data-aug}
\begin{figure}[t]
    \centering
    \includegraphics[width=0.9\linewidth]{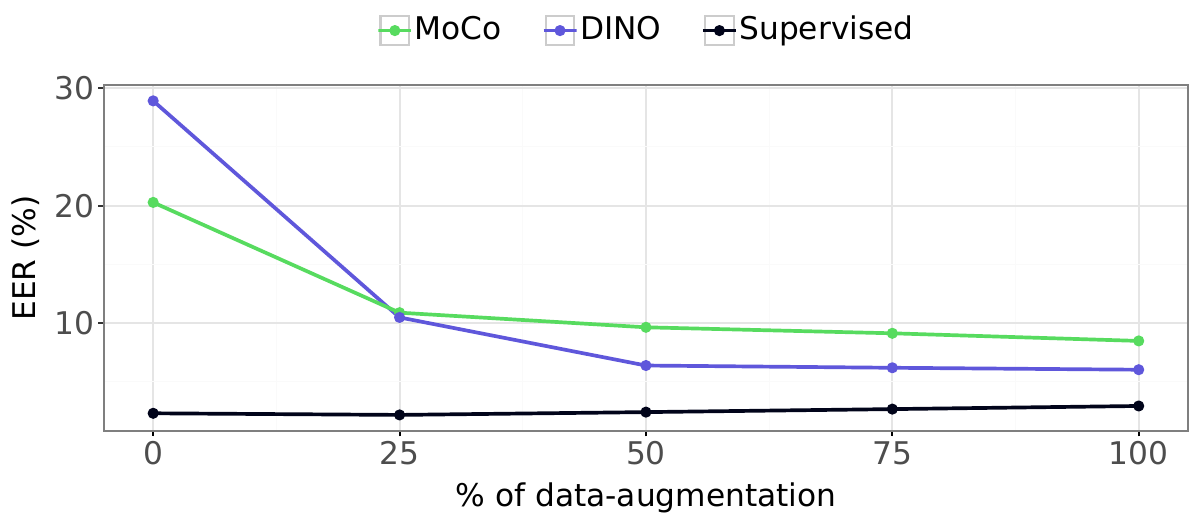}
    \titlecaption{Performance of \ac{SSL} frameworks on \ac{SV} with different probabilities of applying data-augmentation}{The encoder is Fast ResNet-34 and the \acs{EER} is reported on VoxCeleb1-O.}
    \label{fig:data_aug}
\end{figure}
\begin{table}[t]
  \footnotesize
  \titlecaption{Effect of data-augmentation on \ac{SV} performance of \ac{SSL} frameworks}{The encoder is Fast ResNet-34 and the benchmark is VoxCeleb1-O.}
  \label{tab:data_aug}
  \centering
  \begin{tabular}{lcS[table-format=1.2]S[table-format=1.4]}
    \toprule
    \textbf{Framework} & \textbf{Data-augmentation} & \textbf{\eer} & \textbf{\mindcf} \\
    \midrule
    SimCLR & \multirow{6}{*}{\xmark} & 23.11 & 0.8224 \\
    MoCo & & 20.27 & 0.8084 \\
    SwAV & & 30.29 & 0.8285 \\
    VICReg & & 30.02 & 0.8362 \\
    DINO & & 28.91 & 0.8306 \\
    \rowcolor{BaselineRowBg} Supervised & & 2.33 & 0.2092 \\
    \midrule
    SimCLR & \multirow{6}{*}{\cmark} & 9.05 & 0.6364 \\
    MoCo & & 8.49 & 0.5990 \\
    SwAV & & 11.82 & 0.7177 \\
    VICReg & & 11.33 & 0.6658 \\
    DINO & & 6.04 & 0.4526 \\
    \rowcolor{BaselineRowBg} Supervised & & 2.95 & 0.3122 \\
    \bottomrule
  \end{tabular}
\end{table}

Data-augmentation is a fundamental component of SSL frameworks as it determines the quality of the representations. The transformations applied independently to the anchor and the positive samples are designed to approximate the distribution of samples from the same class (i.e., intra-speaker variability) by replicating various acoustic conditions, thereby helping the model learn speaker-discriminative features robust against channel-related characteristics. It is worth noting that the data-augmentation strategy employed in the current setup (e.g., background noise and reverberation) remains relatively simple, in line with common practice in many SSL approaches. This suggests a potential for further improvement through the development of more sophisticated strategies \cite{chen2022ComprehensiveStudySelfDistillation}.

Table~\ref{tab:data_aug} illustrates the impact of applying data-augmentation during SSL training for major SSL frameworks, based on their performance on VoxCeleb1-O. Disabling data-augmentation leads to a substantial degradation in downstream performance, with an average relative increase of 198.8\% in EER. The effect is particularly pronounced for non-contrastive frameworks such as SwAV, VICReg, and DINO. In contrast, the supervised baseline performs better without augmentation, achieving 2.33\% EER and 0.2092 minDCF compared to 2.95\% EER and 0.3122 minDCF with augmentation.

To further assess the dependency of SSL frameworks on data-augmentation, Figure~\ref{fig:data_aug} reports the EER on VoxCeleb1-O as a function of the proportion of data-augmentation applied during training for MoCo and DINO. Both frameworks exhibit a significant performance improvement when the augmentation ratio increases from 0\% to 25\%, with further gains observed as the proportion continues to increase. By comparison, the supervised baseline shows relatively stable performance across different configurations, achieving its best result without data-augmentation.

\subsection{Training distribution}
\label{sec:study:training-dist}
\begin{figure}[t]
    \centering
    \includegraphics[width=0.85\linewidth]{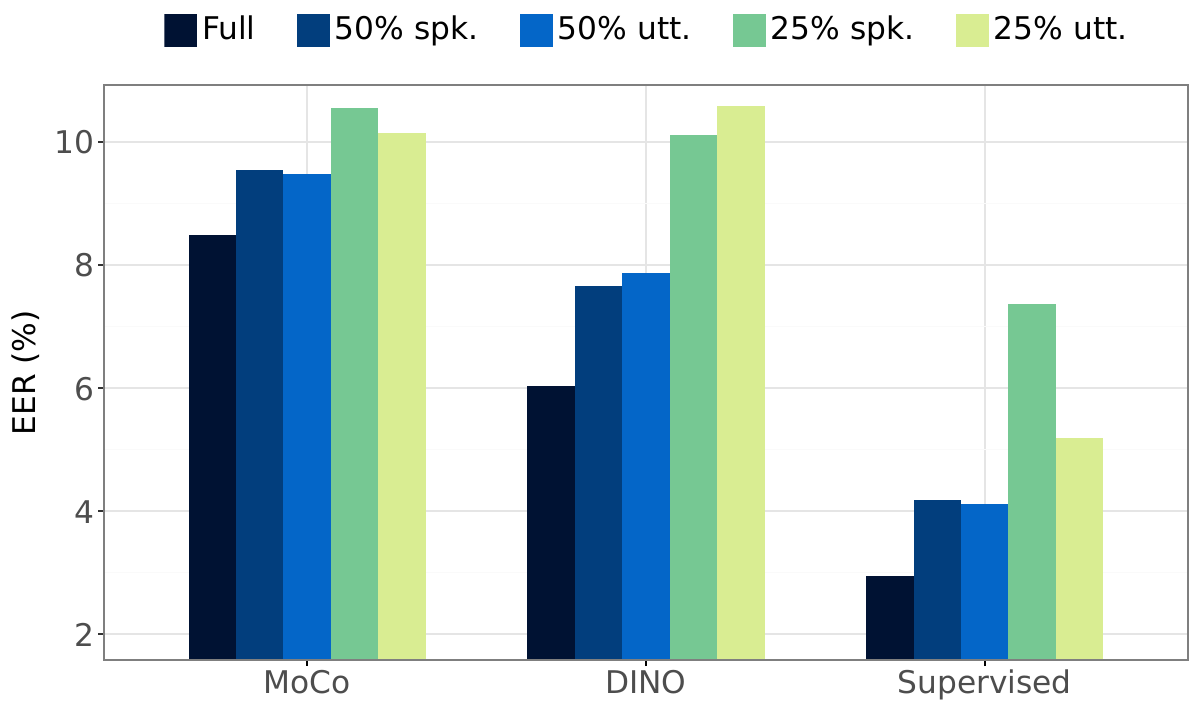}
    \titlecaption{Performance of \ac{SSL} frameworks on \ac{SV} with different distributions of training samples}{The encoder is Fast ResNet-34 and the \acs{EER} is reported on VoxCeleb1-O.}
    \label{fig:training_distribution}
\end{figure}

To gain a deeper understanding of SSL frameworks, the effect of the size and distribution of training data is evaluated. Figure~\ref{fig:training_distribution} presents the EER on VoxCeleb1-O for MoCo, DINO, and the supervised baseline when trained on the full dataset, half of the speakers (50\% spk.), half of the utterances (50\% utt.), a quarter of the speakers (25\% spk.), and a quarter of the utterances (25\% utt.). To conduct this experiment, subsets of the training set are created either by randomly sampling a percentage of speakers and including all their utterances or by directly sampling a percentage of the total utterances.

Regarding the training distribution, both the MoCo framework and the supervised system perform better when more speaker identities are included in the training set (50\% utterances and 25\% utterances), compared to training with a similar number of samples but fewer speaker identities (50\% speakers and 25\% speakers). This outcome aligns with the nature of the underlying losses, notably NT-Xent in Eq.~(\ref{eq:moco_loss}) and AAM-Softmax in Eq.~(\ref{eq:aamsoftmax_loss}), which benefit from numerous speaker identities to model inter-class variability. In contrast, the DINO framework achieves better performance when the training set contains more utterances per speaker (50\% speakers and 25\% speakers), compared to training with more speaker identities and fewer utterances per speaker (50\% utterances and 25\% utterances). This suggests that DINO focuses on capturing intra-speaker variability as its learning task, whereas MoCo and the supervised method primarily learn inter-speaker variability. These results confirm the study presented in \cite{chen2022ComprehensiveStudySelfDistillation} for DINO. This experiment aims to determine whether SSL frameworks benefit more from a larger number of speaker identities or from having more samples per speaker, potentially guiding future unlabeled data collection strategies.

As an additional observation, MoCo demonstrates greater robustness to training data size reduction than both DINO and the supervised baseline, as reflected by a relatively smaller increase in EER when the training data is reduced from 100\% to 50\% and from 100\% to 25\%.

\subsection{Projector}
\label{sec:study:projector}
\begin{table}[t]
  \footnotesize
  \titlecaption{Effect of the projector and positive sampling of \ac{SSL} frameworks on \ac{SV} performance}{The encoder is Fast ResNet-34 and the benchmark is VoxCeleb1-O. \textbf{$\Delta$} corresponds to the difference between the EER without and with a projector.}
  \label{tab:projector_pos_sampling}
  \centering
  \begin{tabular}{lcS[table-format=2.2]S[table-format=2.2]S[table-format=1.2]}
    \toprule
    \multirow{2}{*}{\textbf{Framework}} & \multirow{2}{*}{\textbf{Pos. sampling}} & \textit{Without proj.} & \textit{With proj.} & {\multirow{2}{*}{\textbf{$\Delta$}}} \\
    \cmidrule(lr){3-3} \cmidrule(lr){4-4}
    & & \textbf{\eer} & \textbf{\eer} & \\
    \midrule
    SimCLR & \multirow{5}{*}{SSL} & \bfseries 9.05 & 10.91 & -1.86 \\
    MoCo & & \bfseries 8.49 & 12.46 & -3.97 \\
    SwAV & & 13.07 & \bfseries 11.82 & 1.25 \\
    VICReg & & 20.53 & \bfseries 11.33 & 9.20 \\
    DINO & & 14.23 & \bfseries 6.04 & 8.19 \\
    \midrule
    SimCLR & \multirow{5}{*}{Supervised} & \bfseries 3.66 & 6.19 & -2.53 \\
    MoCo & & \bfseries 3.73 & 6.72 & -2.99 \\
    SwAV & & 9.78 & \bfseries 8.17 & 1.61 \\
    VICReg & & 13.46 & \bfseries 6.63 & 6.83 \\
    DINO & & 8.90 & \bfseries 4.71 & 4.19 \\
    \bottomrule
  \end{tabular}
\end{table}

The projector has a crucial role in SSL frameworks by mapping intermediate representations, used for the downstream task, to embeddings, used for the pretext task (i.e., for loss computation and model optimization). Its purpose is to encourage the model to learn more generalizable representations that are not overly tailored to the pretext task, thereby mitigating potential mismatches between pretext and downstream objectives and reducing overfitting to the biases of the training task \cite{balestriero2023CookbookSSL,bordes2023GuillotineRegularization}. This strategy is widely adopted across other applications, such as in x-vector architectures where intermediate embeddings are extracted before the classification layer \cite{snyder2018XVectors}, or in transfer learning where representations from earlier layers are reused across tasks \cite{yosinski2014Transferable}. In the context of SV, such discrepancies may arise when the SSL objective fails to capture speaker-discriminative information or instead emphasizes irrelevant variations.

The performance of SSL frameworks with and without a projector is reported in the top section of  Table~\ref{tab:projector_pos_sampling} in terms of EER on VoxCeleb1-O, where $\Delta$ denotes the difference between the EER without and with a projector. Contrastive frameworks such as SimCLR and MoCo perform best without a projector ($\Delta < 0$), whereas SwAV, VICReg, and DINO achieve better results with a projector ($\Delta > 0$). These findings suggest that contrastive-based frameworks are more naturally aligned with the downstream task, as their loss functions explicitly encourage discrimination between samples. In contrast, non-contrastive frameworks benefit from the additional abstraction introduced by the projector.

The same experiment is repeated using a supervised positive sampling strategy, where ground-truth labels are used to form anchor-positive pairs. As mentioned in \cite{bordes2023GuillotineRegularization}, this technique enables reducing the mismatch between the pretext and downstream tasks by eliminating the bias caused by same-utterance anchor-positive pairs. The results, shown in the bottom section of Table~\ref{tab:projector_pos_sampling}, demonstrate a performance improvement for all frameworks. For non-contrastive frameworks (such as VICReg and DINO), the performance difference between configurations with and without a projector decreases ($\Delta$ is smaller), indicating a smaller performance gain from its use. This suggests that enhancing the alignment of the training objective with the evaluation task reduces the reliance on a projector. Notably, contrastive frameworks (such as SimCLR and MoCo), which perform better without a projector in the SSL setting, continue to achieve the best performance without one ($\Delta < 0$). However, SwAV, VICReg, and DINO frameworks still perform best with a projector, implying that their objective functions are less effective at capturing speaker-discriminative features and may benefit from a deeper projector architecture.

\subsection{Positive sampling}
\label{sec:study:pos-sampling}
\begin{figure}[t]
    \centering
    \includegraphics[width=0.85\linewidth]{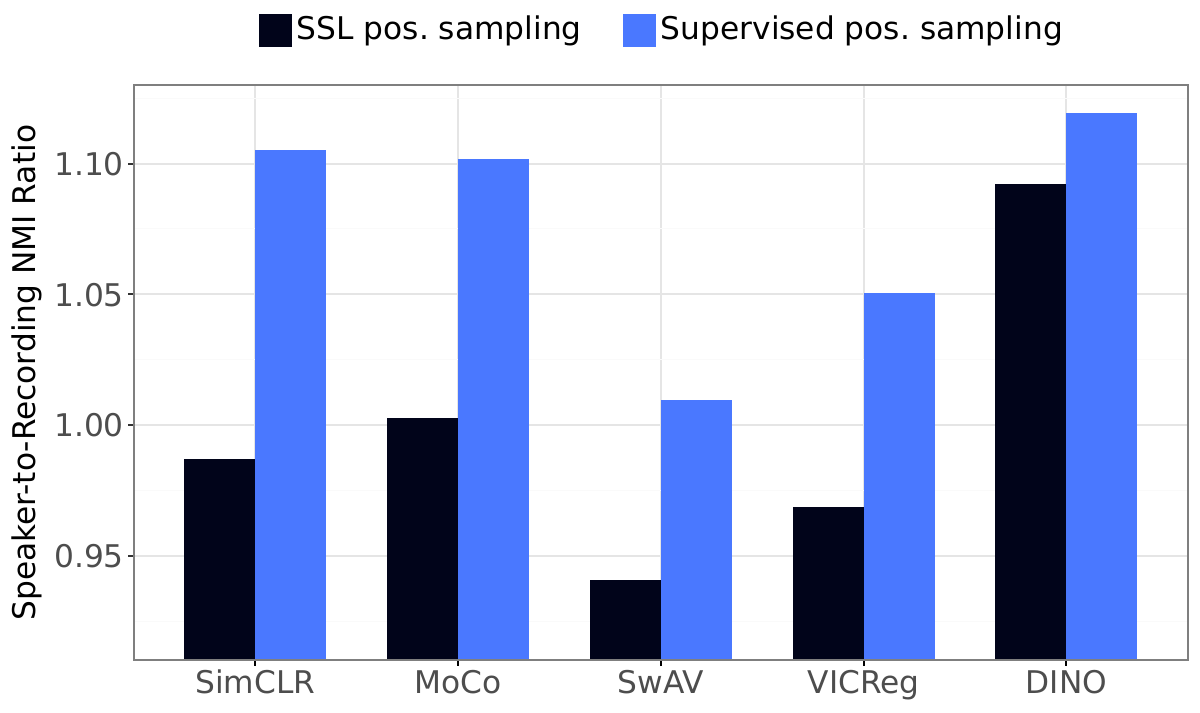}
    \titlecaption{Speaker-to-Recording Ratio (SRR) for \ac{SSL} frameworks with SSL and Supervised positive sampling}{The encoder is Fast ResNet-34, and the \ac{NMI} is computed with k-means clustering assignments (\num{1251} clusters) on representations extracted from VoxCeleb1.}
    \label{fig:nmi_pos_sampling}
\end{figure}

Positive sampling is a fundamental aspect of SSL frameworks since it determines how anchor-positive pairs are constructed. This is essential for guiding the model to learn speaker-discriminative features robust to extrinsic variations by generating relevant anchor-positive pairs that shape the speaker distribution in the latent space, similar to data-augmentation, which serves the same purpose. Several studies have shown that the definition of positive examples is a primary factor influencing the quality of SSL representations \cite{balestriero2023CookbookSSL}.

Since the default SSL positive sampling creates anchor-positive pairs from segments extracted from the same utterance (and thus with the same recording conditions), models are prone to encode channel-related information in the representations. This issue is further amplified given that datasets like VoxCeleb contain real-world recordings affected by a variety of channel effects and other factors. Therefore, the limitations of same-utterance positive sampling are commonly addressed through extensive data-augmentation, which simulates diverse acoustic conditions to mitigate this bias.

As shown in the bottom section of Table~\ref{tab:projector_pos_sampling}, the downstream performance of all SSL frameworks is improved when using VoxCeleb metadata to generate anchor-positive pairs from distinct recordings of the same speaker. This highlights the benefit of using different recordings to form positive pairs, as it encourages the model to associate varied acoustic conditions of the same speaker with a consistent identity, similar to the supervised training scenario. Specifically, this supervised positive sampling results in EER reductions of approximately 60\% and 56\% for SimCLR and MoCo, respectively, on VoxCeleb1-O. These findings demonstrate the negative impact of the default same-utterance positive sampling on the performance of SSL frameworks for SV. Furthermore, this suggests that conventional audio augmentation techniques are insufficient to simulate the diversity among samples from the same class and fully address this limitation.

Let SRR denote the Speaker-to-Recording Ratio, which quantifies the alignment of the learned representations with speaker identity rather than recording characteristics. A higher SRR thus indicates better alignment with the downstream task. SRR is defined as:
\begin{equation}
    \label{eq:srr}
    \text{SRR} = \frac{\text{NMI}(\hat{\mathcal{S}}, \mathcal{S}_{\text{speakers}})}{\text{NMI}(\hat{\mathcal{S}}, \mathcal{S}_{\text{recordings}})},
\end{equation}
where $\hat{\mathcal{S}}$ denotes the k-means clustering assignments of VoxCeleb1 representations (with \num{1251} clusters, matching the number of speaker identities), $\mathcal{S}_{\text{speakers}}$ are speaker labels, and $\mathcal{S}_{\text{recordings}}$ are video labels, both derived from VoxCeleb1 metadata.
The Normalized Mutual Information (NMI) \cite{strehl2002ClusterEnsembles} quantifies the alignment between the predicted cluster assignments $\hat{\mathcal{S}}$ and the ground-truth labels $\mathcal{S}$, and is defined as:
\begin{equation}
    \label{eq:nmi}
    \text{NMI}(\hat{\mathcal{S}}, \mathcal{S}) = \frac{2 \cdot I(\hat{\mathcal{S}}; \mathcal{S})}{H(\hat{\mathcal{S}}) + H(\mathcal{S})},
\end{equation}
where $I(\hat{\mathcal{S}}; \mathcal{S})$ is the Mutual Information (MI) between the two sets, and $H(\hat{\mathcal{S}})$ and $H(\mathcal{S})$ are their respective entropies.

To confirm the previous findings, Figure~\ref{fig:nmi_pos_sampling} reports the Speaker-to-Recording Ratio (SSR) for major SSL frameworks with SSL (black) and Supervised (blue) positive sampling on VoxCeleb1. Compared to SSL positive sampling, employing the supervised positive sampling yields higher SSR values, indicating that information related to recording source (i.e., VoxCeleb videos) is less prevalent in the representations. This results in an average relative improvement of 8\% over the same-utterance positive sampling. Note that DINO is more robust to this bias as it provides the highest ratio among approaches based on the SSL positive sampling. This underscores the role of positive sampling in shaping what information is encoded in the representations.
\section{Evaluation and Comparison on Speaker Verification}
\label{sec:evaluation}

This section provides a comprehensive evaluation of SSL frameworks, considering multiple encoder architectures and benchmarks. The analysis includes both in-domain and out-of-domain scenarios, with performance compared to supervised baselines. Single- and multi-stage SSL methods are reviewed and evaluated on VoxCeleb benchmarks to highlight key trends and developments in the field. Finally, a label-efficient evaluation setting is introduced to underscore the potential of SSL approaches for reducing reliance on labeled data.

\subsection{Evaluation of SSL frameworks}
\label{sec:evaluation:frameworks}
\newcolumntype{H}{>{\setbox0=\hbox\bgroup}c<{\egroup}@{}}

\begin{table*}[t]
  \footnotesize
  \titlecaption{Evaluation of \ac{SSL} frameworks on \ac{SV}}{Performance is reported on all VoxCeleb benchmarks, sorted by \ac{EER} on VoxCeleb1-O $\downarrow$, with Fast ResNet-34 and ECAPA-TDNN encoders. The best method for each encoder is highlighted in \colorbox{SelectedRowBg}{light blue} and the baselines are highlighted in \colorbox{BaselineRowBg}{light gray}. The fusion comprises all SSL approaches trained with the ECAPA-TDNN encoder.}
  \label{tab:ssl_frameworks_sv}
  \centering
  \begin{tabular}{lcS[table-format=2.2]S[table-format=2.2]S[table-format=1.4]S[table-format=2.2]S[table-format=1.4]S[table-format=2.2]S[table-format=1.4]}
    \toprule
    \multirow{2}{*}{\textbf{Framework}} & \multirow{2}{*}{\textbf{Encoder}} & {\multirow{2}{*}{\textbf{FLOPs (G)}}} & \multicolumn{2}{c}{\textbf{VoxCeleb1-O}} & \multicolumn{2}{c}{\textbf{VoxCeleb1-E}} & \multicolumn{2}{c}{\textbf{VoxCeleb1-H}} \\
    \cmidrule(lr){4-5} \cmidrule(lr){6-7} \cmidrule(lr){8-9}
    & & & \textbf{\eer} & \textbf{\mindcf} & \textbf{\eer} & \textbf{\mindcf} & \textbf{\eer} & \textbf{\mindcf} \\
    \midrule
    \rowcolor{BaselineRowBg} Random & & & 42.05 & 0.9969 & 42.38 & 0.9994 & 43.94 & 0.9997 \\
    \midrule
    CPC & \multirow{12}{*}{Fast ResNet-34} & 1.92 & 12.77 & 0.8033 & 13.75 & 0.8526 & 21.47 & 0.9048 \\
    SimCLR & & 1.80 & 9.05 & 0.6364 & 9.79 & 0.6769 & 15.21 & 0.7664 \\
    MoCo & & 3.66 & 8.49 & 0.5990 & 9.21 & 0.6588 & 14.25 & 0.7503 \\
    DeepCluster & & 1.84 & 15.16 & 0.8193 & 16.53 & 0.8941 & 23.40 & 0.9437 \\
    SwAV & & 1.84 & 11.82 & 0.7177 & 13.25 & 0.8301 & 20.19 & 0.9100 \\
    W-MSE & & 1.80 & 14.62 & 0.8506 & 15.98 & 0.9342 & 24.83 & 0.9810 \\
    Barlow Twins & & 1.82 & 13.22 & 0.7658 & 14.31 & 0.8424 & 20.54 & 0.9080 \\
    VICReg & & 1.82 & 11.33 & 0.6658 & 12.75 & 0.7581 & 18.92 & 0.8577 \\
    BYOL & & 3.66 & 13.99 & 0.7509 & 14.94 & 0.8498 & 21.85 & 0.9181 \\
    SimSiam & & 1.84 & 28.94	& 0.9984 & 27.80 & 0.9997 & 41.53 & 0.9997 \\
    \rowcolor{SelectedRowBg} DINO & & 10.90 & \bfseries 6.04 & \bfseries 0.4526 & \bfseries 6.92 & \bfseries 0.5272 & \bfseries 10.54 & \bfseries 0.6456 \\
    \rowcolor{BaselineRowBg} Supervised & & 0.91 & 2.95 & 0.3122 & 3.01 & 0.3475 & 5.45 & 0.4993 \\
    \midrule
    SimCLR & \multirow{6}{*}{ECAPA-TDNN} & 14.94 & 6.41 & 0.5160 & 6.91 & 0.5616 & 11.06 & 0.6708 \\
    MoCo & & 29.95 & 6.48 & 0.5372 & 6.77 & 0.5635 & 11.01 & 0.6700 \\
    SwAV & & 14.98 & 8.12	& 0.6148 & 9.12 & 0.7119 & 15.54 & 0.8282 \\
    VICReg & & 14.97 & 7.42 & 0.5659 & 8.75 & 0.6970 & 15.01 & 0.8518 \\
    \rowcolor{SelectedRowBg} DINO & & 90.87 & \bfseries 2.82 & \bfseries 0.3463 & \bfseries 3.03 & \bfseries 0.3923 & \bfseries 5.93 & \bfseries 0.5756 \\
    \rowcolor{BaselineRowBg} Supervised & & 7.48 & 1.34 & 0.1521 & 1.49 & 0.1736 & 2.84 & 0.2887 \\
    \midrule
    \textit{Fusion} & & & \bfseries 2.60 & \bfseries 0.2886 & \bfseries 2.88 & \bfseries 0.3426 & \bfseries 5.19 & \bfseries 0.4997 \\
    \bottomrule
  \end{tabular}
\end{table*}

\begin{figure}[t]
    \centering
    \includegraphics[width=0.9\linewidth]{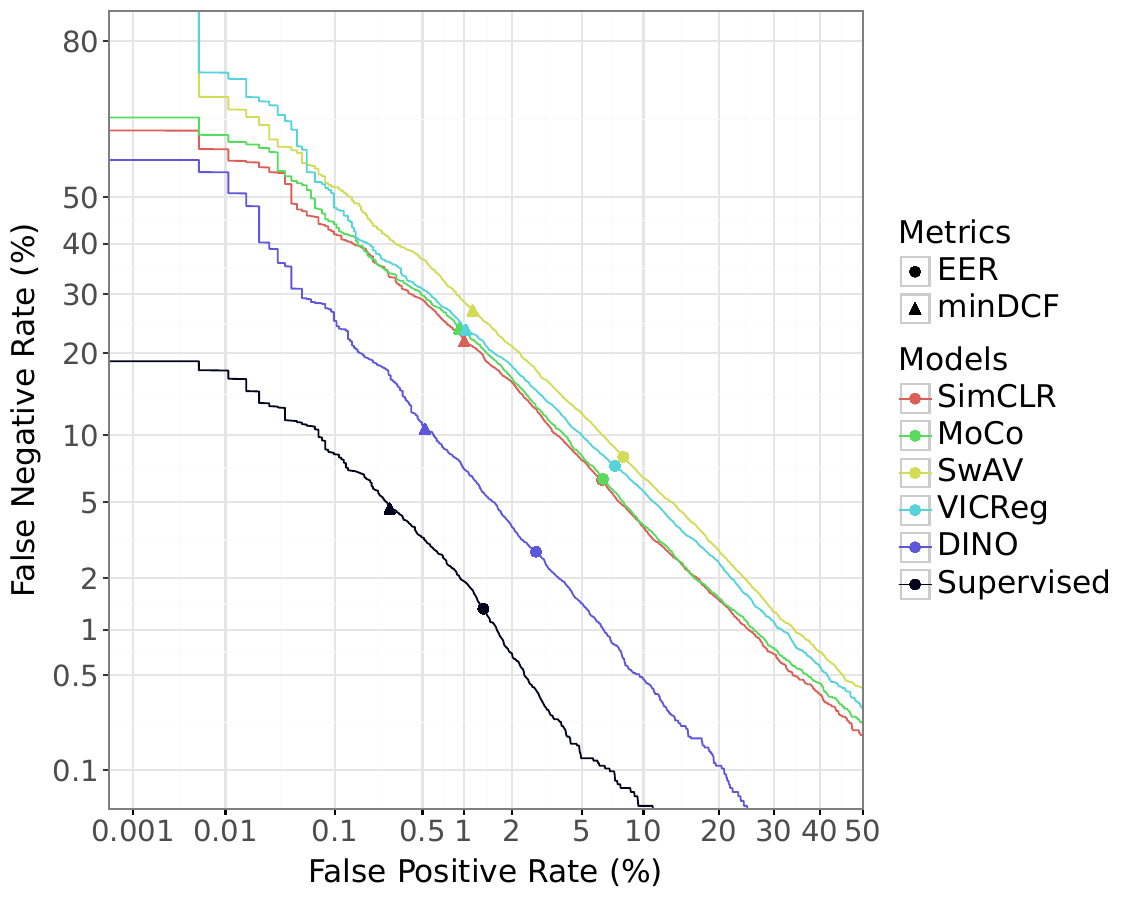}
    \titlecaption{\acf{DET} curves of \ac{SSL} frameworks on \ac{SV}}{The encoder is ECAPA-TDNN and the benchmark is VoxCeleb1-O.}
    \label{fig:det_curve}
\end{figure}
\begin{table}[t]
  \footnotesize
  \titlecaption{Evaluation of \ac{SSL} frameworks on out-of-domain \ac{SV} benchmarks: SITW and VOiCES}{Performance is reported on \texttt{SITW eval core-core} and \texttt{VOiCES dev}. The encoder is ECAPA-TDNN. The best method is highlighted in \colorbox{SelectedRowBg}{light blue} and the supervised baseline is highlighted in \colorbox{BaselineRowBg}{light gray}.}
  \label{tab:ssl_frameworks_sv_out_of_domain}
  \centering
  \begin{tabular}{lS[table-format=2.2]S[table-format=1.4]S[table-format=1.2]S[table-format=1.4]}
    \toprule
    \multirow{2}{*}{\textbf{Framework}} & \multicolumn{2}{c}{\textbf{SITW}} & \multicolumn{2}{c}{\textbf{VOiCES}} \\
    \cmidrule(lr){2-3} \cmidrule(lr){4-5}
    & \textbf{\eer} & \textbf{\mindcf} & \textbf{\eer} & \textbf{\mindcf} \\
    \midrule
    SimCLR & 10.09 & 0.8010 & 3.92 & 0.4172 \\
    MoCo   & 10.01 & 0.7779 & 3.96 & 0.4212 \\
    SwAV   & 13.68 & 0.8709 & 5.79 & 0.5954 \\
    VICReg & 13.59 & 0.8453 & 5.59 & 0.5798 \\
    \rowcolor{SelectedRowBg} DINO   & \bfseries 4.92 & \bfseries 0.5123 & \bfseries 1.85 & \bfseries 0.2656 \\
    \rowcolor{BaselineRowBg} Supervised & 2.05 & 0.2032 & 1.15 & 0.1437 \\
    \bottomrule
  \end{tabular}
\end{table}

\begin{figure}[t]
    \centering
    \includegraphics[width=0.65\linewidth]{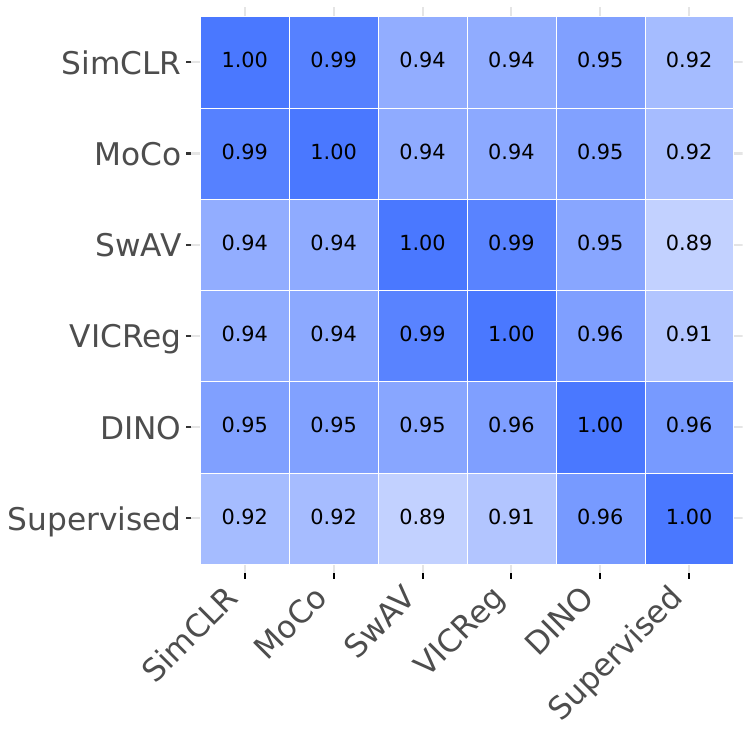}
    \titlecaption{Correlation of scores generated by \ac{SSL} frameworks on \ac{SV}}{The encoder is ECAPA-TDNN and the benchmark is VoxCeleb1-O.}
    \label{fig:correlation}
\end{figure}

A comprehensive comparison of SSL frameworks for SV on all VoxCeleb1 benchmarks is presented in Table~\ref{tab:ssl_frameworks_sv}. Two different encoder architectures are considered, namely the lightweight Fast ResNet-34 (top section) and the larger ECAPA-TDNN (bottom section).

Using the Fast ResNet-34 encoder, DINO achieves the lowest EER on VoxCeleb1-O at 6.04\%, followed by MoCo (8.49\% EER), SimCLR (9.05\% EER), VICReg (11.33\% EER), and SwAV (11.82\% EER). For comparison, the supervised baseline reaches 2.95\% EER on the same benchmark. These results highlight the effectiveness of the DINO framework, although SimCLR and MoCo remain competitive while requiring less computational complexity (\SI{10.90}{\giga\flop} compared to \SI{1.80}{\giga\flop} and \SI{3.66}{\giga\flop}, respectively). In contrast, W-MSE, DeepCluster, and SimSiam perform significantly worse, indicating a limited ability to learn discriminative speaker representations without supervision. Notably, SimSiam fails to prevent collapse, in contrast to BYOL, which incorporates an EMA update between student and teacher networks as an effective collapse-prevention mechanism. The following focuses on the five best-performing SSL frameworks, covering a diverse set of training paradigms: SimCLR and MoCo (\textit{contrastive learning}), SwAV (\textit{clustering}), VICReg (\textit{information maximization}), and DINO (\textit{self-distillation}).

Switching to the ECAPA-TDNN encoder, a substantial improvement is observed across all frameworks. DINO achieves the best performance, with an EER of 2.82\% on VoxCeleb1-O, narrowing the gap with the supervised baseline at 1.34\% EER. SimCLR and MoCo achieve 6.41\% and 6.48\% EER, respectively, followed by VICReg (7.42\% EER) and SwAV (8.12\% EER). These results suggest that SSL frameworks benefit significantly from increased encoder capacity, as expected and consistent with prior observations in other domains. Similar trends and relative performance differences can be observed on the VoxCeleb1-E and VoxCeleb1-H trials. The DET curves shown in Figure~\ref{fig:det_curve} further illustrate the performance differences between SSL frameworks on the VoxCeleb1-O trial set, confirming the superiority of DINO in terms of detection error trade-off. The combination of these systems results in 2.60\% EER on VoxCeleb1-O, as indicated by the last row of Table~\ref{tab:ssl_frameworks_sv}. While the correlation matrix in Figure~\ref{fig:correlation} shows strong similarities between the systems, the fusion still yields measurable gains, suggesting that the frameworks provide limited but non-negligible complementarity. Despite the promising improvements from SSL, supervised pre-training still yields the best performance, highlighting the continued relevance of supervised learning. As reference points, a random baseline yields an EER above 42\% across all trials, underscoring the difficulty of the SV task in the absence of meaningful representation learning. Furthermore, human performance, as reported in \cite{huh2020AAT}, reaches 26.51\% EER for crowd-sourced annotators (Amazon Mechanical Turk) and 15.77\% EER for a group of SR researchers, highlighting the challenge even for humans under these conditions.

The performance gap between DINO and alternative approaches becomes more pronounced with ECAPA-TDNN, indicating that the effectiveness of its design scales better with encoder complexity. However, SimCLR and MoCo stand out due to their competitive results combined with lower computational requirements. This observation supports the hypothesis that contrastive-based objectives, such as those used in SimCLR and MoCo, are particularly effective for SV.

Finally, Table~\ref{tab:ssl_frameworks_sv_out_of_domain} reports results on out-of-domain settings using SITW eval. core-core \cite{mclaren2016SITW} and VOiCES dev. \cite{nandwana2019VOiCES} benchmarks. Similarly to previous findings, DINO consistently outperforms other SSL frameworks by a considerable margin, further demonstrating its strong generalization capabilities across diverse conditions. SimCLR, MoCo, SwAV, and VICReg yield moderate performance under challenging conditions, preserving the relative performance differences observed on VoxCeleb.

\subsection{Comparison of single and multi-stage SSL methods}
\label{sec:evaluation:methods}
\definecolor{dcf001}{rgb}{0.0, 0.0, 0.0}
\definecolor{dcf005}{rgb}{0.6, 0.6, 0.6}

\newcommand{\mindcfall}{$\text{minDCF}_\text{\textcolor{dcf001}{0.01}/\textcolor{dcf005}{0.05}}$}

\begin{table*}[t]
  \scriptsize
  \titlecaption{Comparison of single-stage \ac{SSL} methods on \ac{SV}}{Results are reported on all VoxCeleb benchmarks, sorted by EER on VoxCeleb1-O $\downarrow$. To provide a complete overview, both $\text{minDCF}_\text{0.01}$ (black) and $\text{minDCF}_\text{0.05}$ (\textcolor{dcf005}{light gray}) are reported, using $P_{target}=0.01$ and $P_{target}=0.05$, respectively. Methods relying on other modalities for training are indicated with an asterisk (*), and only methods using speech data for evaluation are considered. Cells marked with a gray dot (\m{}) indicate missing values.}
  \label{tab:ssl_methods_sv_single-stage}
  \centering
  \begin{tabular}{lllS[table-format=2.2]S[table-format=1.4]S[table-format=1.2]S[table-format=1.4]S[table-format=2.2]S[table-format=1.4]}
    \toprule
    \multirow{2}{*}{\textbf{Category}} & \multirow{2}{*}{\textbf{Method}} & \multirow{2}{*}{\textbf{Encoder}} & \multicolumn{2}{c}{\textbf{VoxCeleb1-O}} & \multicolumn{2}{c}{\textbf{VoxCeleb1-E}} & \multicolumn{2}{c}{\textbf{VoxCeleb1-H}} \\
    \cmidrule(lr){4-5} \cmidrule(lr){6-7} \cmidrule(lr){8-9}
    & & & \textbf{\eer} & \textbf{\mindcfall} & \textbf{\eer} & \textbf{\mindcfall} & \textbf{\eer} & \textbf{\mindcfall} \\
    \midrule

    \multirow{16}{*}{\shortstack[l]{Contrastive}} & Disent.* \cite{nagrani2020Disentangled} & VGG-M-40 & 22.09 & \m & \m & \m & \m & \m \\
    & CDDL* \cite{chung2020CDDL} & VGG-M-40 & 17.52 & \m & \m & \m & \m & \m \\
    & GCL \cite{inoue2020GCL} & ResNet-18 & 15.26 & \m & \m & \m & \m & \m \\
    & AP + i-mix \cite{kang2022LMix} & ECAPA-TDNN & 10.63 & \m & \m & \m & \m & \m \\
    & AP + l-mix \cite{kang2022LMix} & ECAPA-TDNN & 10.49 & \m & \m & \m & \m & \m \\
    & AP + AAT \cite{huh2020AAT} & Fast ResNet-34 & 8.65 & \textcolor{dcf005}{0.4540} & \m & \m & \m & \m \\
    & Contrastive + VICReg \cite{lepage2022LabelEfficient} & Thin ResNet-34 & 8.47 & 0.6400 & \m & \m & \m & \m \\
    & SimCLR + MSE loss \cite{zhang2021SimCLR} & Thin ResNet-34 & 8.28 & 0.6100 & \m & \m & \m & \m \\
    & MoCo + ProtoNCE \cite{xia2021MoCo} & TDNN & 8.23 & 0.5900 & \m & \m & \m & \m \\
    & CEL \cite{mun2020CEL} & Fast ResNet-34 & 8.01 & \m & \m & \m & \m & \m \\
    & SimCLR + Margin \cite{lepage2024AdditiveMargin} & Fast ResNet-34 & 7.85 & 0.6168 & \m & \m & \m & \m \\
    & SimCLR + AAT \cite{tao2022LGL} & ECAPA-TDNN & 7.36 & \m & 7.90 & \m & 12.32 & \m \\
    & W-ACSG \cite{gan2024WACSG} & ECAPA-TDNN & 7.02 & \textcolor{dcf005}{0.3900} & \m & \m & \m & \m \\
    & C3-MoCo \cite{zhang2022C3DINO} & ECAPA-TDNN & 6.40 & \m & \m & \m & \m & \m \\
    & DPP* \cite{tao2023DPP} & ECAPA-TDNN & 2.89 & \m & 3.17 & \m & 6.27 & \m \\
    & SimCLR-SSPS \cite{lepage2025SSPS} & ECAPA-TDNN & \bfseries 2.57 & \bfseries 0.3033 & \bfseries 3.11 & \bfseries 0.3125 & \bfseries 5.56 & \bfseries 0.4638 \\
    \midrule
    
    \multirow{16}{*}{\shortstack[l]{Self-distillation}} & SSReg \cite{sang2022SSReg} & Thin ResNet-34 & 6.99 & \textcolor{dcf005}{0.4340} & \m & \m & \m & \m \\
    & DINO + Cosine loss \cite{han2022DLGLC} & Thin-ResNet34 & 6.16 & 0.5240 & \m & \m & \m & \m \\
    & DINO \cite{jung2022RawNet3} & RawNet3 & 5.40 & \textcolor{dcf005}{0.3396} & \m & \m & \m & \m \\
    & DINO \cite{cho2022DINO} & Fast ResNet-34 & 4.83 & 0.4630 & \m & \m & \m & \m \\
    & DINO + Curriculum \cite{heo2022DINOCurriculum} & ECAPA-TDNN & 4.47 & \textcolor{dcf005}{0.3057} & \m & \m & \m & \m \\
    & CA-DINO \cite{han2024CADINO} & ECAPA-TDNN & 3.59 & 0.3529 & 3.85 & 0.4182 & 6.92 & 0.5743 \\
    & RDINO \cite{chen2023RDINO} & ECAPA-TDNN & 3.29 & \textcolor{dcf005}{0.2470} & \m & \m & \m & \m \\
    & MeMo \cite{jin2024MeMo} & ECAPA-TDNN & 3.10 & \textcolor{dcf005}{0.2290} & 3.53 & \textcolor{dcf005}{0.2970} & 7.04 & \textcolor{dcf005}{0.5690} \\
    & EBCA-DINO \cite{hao2025EBCADINO} & ECAPA-TDNN & 2.94 & \m & 2.99 & \m & 5.14 & \m \\
    & RDINO + W-GVKT \cite{jin2024WGVKT} & ECAPA-TDNN & 2.89 & 0.3330 & \m & \m & \m & \m \\
    & PDC-RDINO \cite{zhao2024PrototypeDivisionDINO} & ECAPA-TDNN & 2.80 & 0.3150 & \m & \m & \m & \m \\
    & DINO + RMP \cite{kim2024RMP} & BA-Transformer & 2.62 & \textcolor{dcf005}{0.1736} & 2.84 & \textcolor{dcf005}{0.1936} & 5.14 & \textcolor{dcf005}{0.3036} \\
    & DINO-SSPS \cite{lepage2025SSPS} & ECAPA-TDNN & 2.53 & 0.2843 & 2.55 & 0.3150 & 4.93 & 0.4632 \\
    & DINO + Aug. \cite{chen2022ComprehensiveStudySelfDistillation} & ECAPA-TDNN & 2.51 & \textcolor{dcf005}{0.1626} & 2.47 & \m & 4.79 & \m \\
    & C3-DINO \cite{zhang2022C3DINO} & ECAPA-TDNN & 2.50 & \m & \m & \m & \m & \m \\
    & SDPN \cite{chen2025SDPN} & ECAPA-TDNN & \bfseries 1.80 & \bfseries \textcolor{dcf005}{0.1390} & \bfseries 1.99 & \bfseries \textcolor{dcf005}{0.1310} & \bfseries 3.62 & \bfseries \textcolor{dcf005}{0.2190} \\
    \bottomrule
  \end{tabular}
\end{table*}

\begin{table*}[t]
  \scriptsize
  \titlecaption{Comparison of multi-stage \ac{SSL} methods on \ac{SV}}{Results are reported on all VoxCeleb benchmarks, sorted by EER on VoxCeleb1-O $\downarrow$. ``K-M'' refers to k-means and ``AHC'' to Agglomerative Hierarchical Clustering. The ``LF/LC'' column indicates whether methods use pseudo-label filtering (LF) or correction (LC). Methods relying on other modalities for training are indicated with an asterisk (*) and only methods using speech data for evaluation are considered. Cells marked with a gray dot (\m{}) indicate missing values.}
  \label{tab:ssl_methods_sv_multi-stage}
  \centering
  \begin{tabular}{llllccS[table-format=1.2]S[table-format=1.2]S[table-format=1.2]}
    \toprule
    \multirow{2}{*}{\textbf{Method}} & \multirow{2}{*}{\textbf{Encoder}} & \multirow{2}{*}{\textbf{Base model}} & \multirow{2}{*}{\textbf{Clustering}} & {\multirow{2}{*}{\textbf{\# iters.}}} & {\multirow{2}{*}{\textbf{LF/LC}}} & \textbf{VoxCeleb1-O} & \textbf{VoxCeleb1-E} & \textbf{VoxCeleb1-H} \\
    \cmidrule(lr){7-7} \cmidrule(lr){8-8} \cmidrule(lr){9-9}
    & & & & & & \textbf{\eer} & \textbf{\eer} & \textbf{\eer} \\
    \midrule

    DKU-DukeECE \textit{\tiny(VoxSRC-20)} \cite{wang2020DKUDukeECEVoxSRC20} & Fast ResNet-34 & SimCLR & K-M (6k) & 2 & \cmark & 5.42 & 6.22 & 9.60 \\
    DukeECE \cite{cai2021IterativeFramework} & Fast ResNet-34 & SimCLR & K-M (6k) & 5 & \cmark & 3.45 & 4.02 & 6.57 \\
    CAMSAT \cite{fathan2024CAMSAT} & ECAPA-TDNN & i-vector & CAMSAT & - & \xmark & 3.06 & \m & \m \\
    AdaptiveDrop \cite{fathan2025AdaptiveDrop} & ECAPA-TDNN & i-vector & CAMSAT & - & \cmark & 2.41 & \m & \m \\
    IDLAB \textit{\tiny(VoxSRC-20)} \cite{thienpondt2020IDLABVoxSRC20} & ECAPA-TDNN & MoCo & K-M (50k); AHC (7.5k) & 7 & \xmark & 2.10 & \m & \m \\
    JHU \textit{\tiny(VoxSRC-21)} \cite{cho2021JHUVoxSRC21} & Res2Net-50 & DINO & K-M (50k); AHC (7.5k) & 4 & \xmark & 1.89 & \m & \m \\
    DKU-DukeECE* \textit{\tiny(VoxSRC-21)} \cite{cai2021DKUDukeECEVoxSRC21,cai2022IncorporatingVisualInformation} & Fast ResNet-34 & SimCLR & K-M (6k) & 4 & \cmark & 1.81 & 2.06 & 3.80 \\
    SNU-HIL \textit{\tiny(VoxSRC-21)} \cite{mun2021SNUHILVoxSRC21} & ECAPA-TDNN & CEL & K-M (50k); AHC (7.5k) & 8 & \xmark & 1.66 & \m & \m \\
    LGL \cite{tao2022LGL} & ECAPA-TDNN & SimCLR & K-M (6k) & 5 & \cmark & 1.66 & 2.18 & 3.76 \\
    DLG-LC \cite{han2022DLGLC} & ECAPA-TDNN & DINO & K-M (7.5k) & 5 & \cmark & 1.47 & 1.78 & 3.19 \\
    DPP* \cite{tao2023DPP} & ECAPA-TDNN & SimCLR & K-M (6k) & 4 & \xmark & 1.44 & 1.77 & 3.27 \\
    AT + HT \cite{zhou2024ATHT} & ECAPA-TDNN & DINO & K-M (6k) & 5 & \cmark & 1.35 & \m & \m \\
    BDS-BPLC \cite{wang2025BDSBPLC} & ECAPA-TDNN & DINO & K-M (7.5k) & 5 & \cmark & 1.33 & 1.56 & 2.78 \\
    CA-DINO + DLG-LC* \cite{han2024CADINO} & ECAPA-TDNN & CA-DINO & K-M (7.5k) & 2 & \cmark & 1.29 & 1.57 & 2.80 \\
    Co-Meta* \cite{chen2023SSLAudioVisualCoMeta} & ECAPA-TDNN & SimCLR & K-M (6k) & 7 & \xmark & 1.27 & 1.82 & 3.31 \\
    Sub-PTM \cite{chen2023SubPTM} & WavLM large & WavLM & Infomap & 5 & \cmark & 1.25 & \m & \m \\
    IPL \cite{aldeneh2025IPL} & MFA-Conformer & i-vector & K-M (25k); AHC (7.5k) & 8 & \xmark & 1.06 & 2.16 & 3.61 \\
    SSRL \cite{cai2025SelfSupervisedReflective} & WavLM large & DINO & K-M (8k) & 1 & \cmark & 1.04 & \m & \m \\
    DINO-WavLM \cite{miara2024WavLMSSLSV} & WavLM base+ & DINO & K-M (50k); AHC (7.5k) & 3 & \cmark & \bfseries 0.99 & \bfseries 1.21 & \bfseries 2.35 \\
    \bottomrule
  \end{tabular}
\end{table*}

A comparison of single-stage SSL methods is provided in Table~\ref{tab:ssl_methods_sv_single-stage}, which details the framework category (i.e., contrastive or self-distillation), encoder architecture, and corresponding SV performance on all VoxCeleb benchmarks, sorted by decreasing EER on VoxCeleb1-O. Note that while Disent. \cite{nagrani2020Disentangled}, CDDL \cite{chung2020CDDL}, and DPP \cite{tao2023DPP} rely on additional modalities during training (e.g., faces from videos), only results obtained with speech as input data are reported for a fair comparison.

The encoder strongly influences downstream performance, as it determines the quality and discriminative power of the learned speaker representations. Therefore, differences in encoder architectures and their implementation should be carefully considered when comparing SSL methods. The following are the most common architectures employed in the literature: TDNN-based models \cite{snyder2018XVectors}, which rely on 1D convolutions over temporal features; ResNet-based models \cite{chung2020DefenceMetricLearningSR,chung2020DelvingVoxCeleb}, which use 2D convolutions over spectrograms, residual connections, and Squeeze-and-Excitation (SE) modules \cite{hu2018SqueezeExcitation}; ECAPA-TDNN \cite{desplanques2020ECAPATDNN}, an enhanced TDNN variant with multi-scale feature extraction based on the Res2Net architecture \cite{gao2021Res2Net}; and WavLM \cite{chen2022WavLM}, a transformer-based model pre-trained in a self-supervised manner on large amounts of unlabeled speech data.

Among contrastive-based methods, a clear trend of performance improvement over time can be observed, primarily driven by the introduction of regularization components into the objective function and the adoption of more powerful encoders. Early approaches, such as Disent. \cite{nagrani2020Disentangled}, CDDL \cite{chung2020CDDL}, and GCL \cite{inoue2020GCL}, exhibit significantly higher EERs, largely due to limited encoder capacity. A notable performance improvement is achieved with the adoption of ResNet-34 variants as encoders (i.e., ``Fast'' and ``Thin''). Several approaches introduced regularization strategies to improve representation learning \cite{mun2020CEL,zhang2021SimCLR,lepage2022LabelEfficient}, reduce the encoding of channel-information \cite{huh2020AAT,tao2022LGL}, incorporate margins in the contrastive loss \cite{lepage2024AdditiveMargin}, and address class collisions \cite{xia2021MoCo,zhang2022C3DINO} or hard-negatives \cite{gan2024WACSG}. Methods such as AP + AAT \cite{huh2020AAT}, SimCLR \cite{zhang2021SimCLR}, MoCo \cite{xia2021MoCo}, and CEL \cite{mun2020CEL} achieve 8.65\%, 8.28\%, 8.23\%, and 8.01\% EER on VoxCeleb1-O, respectively, and have since served as strong baselines in the field. More recent methods leveraging the ECAPA-TDNN encoder demonstrate even better downstream performance. At the time of writing, DPP \cite{tao2023DPP} and SimCLR-SSPS \cite{lepage2025SSPS} achieve state-of-the-art results with 2.89\% and 2.57\% EER on VoxCeleb1-O, respectively, using sophisticated positive sampling strategies.

Self-distillation methods demonstrate superior SV performance compared to contrastive-based approaches. Among them, a clear trend favors the DINO framework, which has been adopted by nearly all methods, except SSReg \cite{sang2022SSReg}, which is based on SimSiam. Various approaches have emerged in this line of work: combining contrastive learning and self-distillation frameworks \cite{sang2022SSReg,zhang2022C3DINO}, applying this approach to different encoders \cite{cho2022DINO,jung2022RawNet3}, introducing regularization components \cite{han2022DLGLC,chen2023RDINO}, revisiting the procedure to construct prototypes \cite{zhao2024PrototypeDivisionDINO,chen2025SDPN}, integrating curriculum learning \cite{heo2022DINOCurriculum}, exploring novel positive sampling strategies \cite{han2024CADINO,lepage2025SSPS}, studying the effect of data-augmentation and diversifying input features \cite{chen2022ComprehensiveStudySelfDistillation,jin2024WGVKT}, modifying the self-distillation architecture \cite{jin2024MeMo,hao2025EBCADINO}, and incorporating a masked modeling objective \cite{kim2024RMP}. The ECAPA-TDNN encoder remains the most common backbone, largely due to its recency and effectiveness, though newer architectures like RawNet3 \cite{jung2022RawNet3} and BA-Transformer \cite{kim2024RMP} have also been explored. Several methods have become prominent baselines in the field, including SSReg \cite{sang2022SSReg}, DINO \cite{cho2022DINO}, CA-DINO \cite{han2024CADINO}, RDINO \cite{chen2023RDINO}, and C3-DINO \cite{zhang2022C3DINO}, achieving 6.99\%, 4.83\%, 3.59\%, 3.29\%, and 2.50\% EER on VoxCeleb1-O, respectively. Note that results can vary considerably within methods using the same encoder, as even subtle adjustments to the training setup may cause notable relative differences, as demonstrated in the previous section. At the time of writing, C3-DINO \cite{zhang2022C3DINO} and SDPN \cite{chen2025SDPN} achieve state-of-the-art performance with 2.50\% and 1.80\% EER on VoxCeleb1-O, respectively.

Multi-stage methods have emerged as a natural extension of single-stage methods, aiming to improve model representations using iteratively refined pseudo-labels. Table~\ref{tab:ssl_methods_sv_multi-stage} provides a comparative overview of recent multi-stage methods for SV, detailing the underlying encoders, base models, clustering strategies, number of iterations, whether some form of pseudo-label filtering or correction is applied, and their performance on the VoxCeleb benchmarks. Note that while DKU-DukeECE \textit{\footnotesize(VoxSRC-21)} \cite{cai2021DKUDukeECEVoxSRC21}, DPP \cite{tao2023DPP}, CA-DINO + DLG-LC \cite{han2024CADINO}, and Co-Meta \cite{chen2023SSLAudioVisualCoMeta} rely on additional modalities during training (e.g., faces from videos), only results obtained with speech as input data are reported for fair comparison.

A common strategy is to discard or down-weight unreliable samples at each stage, either based on clustering confidence \cite{wang2020DKUDukeECEVoxSRC20}, loss-based heuristics \cite{tao2022LGL}, or dynamic filtering using training losses \cite{han2022DLGLC}. Several methods incorporate Large Margin Fine-Tuning (LMFT) during the final iterations to further boost performance \cite{thienpondt2020IDLABVoxSRC20,cho2021JHUVoxSRC21}. Other works explore joint representations across modalities \cite{cai2021DKUDukeECEVoxSRC21,tao2023DPP,han2024CADINO,chen2023SSLAudioVisualCoMeta} or hierarchical training schemes to reduce the sensitivity of top layers to noisy labels \cite{zhou2024ATHT}. The trend observed across the methods reveals a preference for more sophisticated encoders, coupled with state-of-the-art single-stage methods as base models. Clustering strategies typically involve either k-means with \num{6000} or \num{7500} clusters, or k-means with \num{50000} clusters followed by AHC with \num{7500} clusters. On average, these methods run for approximately 5 iterations of pseudo-labeling. Some of these methods have emerged as key baselines, including JHU \textit{\footnotesize(VoxSRC-21)} \cite{cho2021JHUVoxSRC21}, DKU-DukeECE \textit{\footnotesize(VoxSRC-21)} \cite{cai2021DKUDukeECEVoxSRC21}, SNU-HIL \textit{\footnotesize(VoxSRC-21)} \cite{mun2021SNUHILVoxSRC21}, LGL \cite{tao2022LGL}, and DLG-LC \cite{han2022DLGLC}, with performance results of 1.89\%, 1.81\%, 1.66\%, 1.66\%, and 1.47\% EER on VoxCeleb1-O, respectively. Notably, several methods were specifically proposed for the VoxSRC-20 \cite{wang2020DKUDukeECEVoxSRC20,thienpondt2020IDLABVoxSRC20} and VoxSRC-21 \cite{cho2021JHUVoxSRC21,cai2021DKUDukeECEVoxSRC21,mun2021SNUHILVoxSRC21} self-supervised challenges. In particular, the following approaches stand out: DKU-DukeECE \textit{\footnotesize(VoxSRC-21)} \cite{cai2021DKUDukeECEVoxSRC21} delivers competitive results with a relatively lightweight encoder, IPL \cite{aldeneh2025IPL} surpasses more complex methods by using i-vectors to generate the initial pseudo-labels, SSRL \cite{cai2025SelfSupervisedReflective} and DINO-WavLM \cite{miara2024WavLMSSLSV} achieve the best performance by leveraging SSL-based speech foundation models. While these multi-stage procedures are computationally expensive, they have consistently improved downstream performance on VoxCeleb benchmarks, demonstrating the effectiveness of using pseudo-labels to fine-tune SSL models across multiple stages. At the time of writing, IPL \cite{aldeneh2025IPL}, SSRL \cite{cai2025SelfSupervisedReflective}, and DINO-WavLM \cite{miara2024WavLMSSLSV} achieve state-of-the-art performance with 1.06\%, 1.04\% and 0.99\% EER on VoxCeleb1-O, respectively.

\subsection{Label-efficient evaluation}
\label{sec:evaluation:label-efficient}
\begin{figure}[t]
    \centering
    \includegraphics[width=0.7\linewidth]{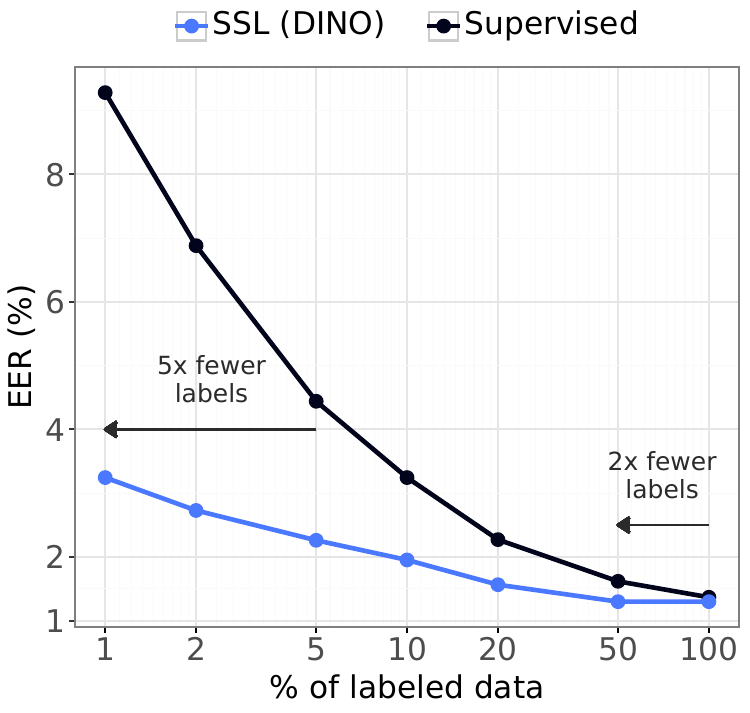}
    \titlecaption{Evaluation of \ac{SSL} fine-tuned with different percentage of labeled data on \ac{SV}}{The model is DINO, the encoder is ECAPA-TDNN and the \ac{EER} is reported on VoxCeleb1-O.}
    \label{fig:label_efficient}
\end{figure}
To assess the label-efficiency of SSL-based models, the EER on VoxCeleb1-O is reported in Figure~\ref{fig:label_efficient}, for both DINO (blue) and the supervised baseline (black), as a function of the proportion of labeled data provided during training. The experiment is conducted using subsets of the VoxCeleb2 training set, with percentages of labeled data varying from 1\% to 100\% out of \num{1092009} samples. The supervised baseline is initialized randomly, while the SSL model is initialized with DINO pre-trained weights and fine-tuned using the AAM-Softmax loss. The results demonstrate that the SSL approach maintains competitive performance even when the amount of labeled data is significantly reduced and that it consistently outperforms the supervised system for all configurations. Notably, the SSL model exhibits performance comparable to that of the supervised counterpart when trained with 2$\times$ (100\% vs. 50\%) or even 5$\times$ (5\% vs. 1\%) fewer labels, as indicated by horizontal arrows. This experiment highlights the strong label-efficiency of SSL models and underscores their potential for low-resource scenarios. By leveraging large amounts of unlabeled speech, SSL reduces the reliance on manual annotation, reaffirming its initial promise as a scalable and efficient approach to SV.
\section{Conclusions \& Perspectives}
\label{sec:conclusions}

This work consists of a review and study of the application of Self-Supervised Learning (SSL) to the downstream task of Speaker Verification (SV).

First, the main SSL frameworks from different paradigms (i.e., contrastive learning, clustering, information maximization, and self-distillation) are presented and applied to SV. A consistent framework for training and evaluation is proposed, ensuring that the performance of each system is assessed under identical conditions. Single-stage and multi-stage SSL methods proposed in the literature and derived from these frameworks are also reviewed.

Subsequently, the effect of the key hyperparameters of SSL frameworks is assessed. In particular, SSL frameworks are very sensitive to hyperparameters, and some of them are fundamental to avoid a collapsing solution. The findings demonstrate that the impact of false negatives (class collisions) in the contrastive loss is negligible in MoCo, with performance improving as the number of negatives increases but deteriorating when it becomes too large. In DINO, having multiple views with different temporal lengths is beneficial, and proper tuning of the teacher temperature is crucial.

Then, a study of SSL training components is conducted to provide more insight into the functioning of these approaches when applied to SV. Notably, the different types of collapse occurring in MoCo and DINO, along with their underlying causes, are identified. The importance of extensive data-augmentation is highlighted, as it plays a crucial role in generating appropriate anchor-positive pairs and simulating various channel conditions. Moreover, for a given number of training samples, DINO is shown to benefit from a smaller number of classes with higher intra-class variability, in contrast to MoCo and the supervised baseline. In addition, the critical role of the projector in mitigating the misalignment between the pretext and downstream task objectives is demonstrated. Most importantly, the positive sampling strategy is found to be fundamental as it enables SSL to model the speaker distribution effectively and encode information that is most relevant to the downstream task. Specifically, the analysis highlights a clear distinction between contrastive and self-distillation paradigms: contrastive methods (e.g., MoCo) predominantly encode inter-speaker variability and exhibit greater robustness to collapse and hyperparameter settings, whereas self-distillation frameworks (e.g., DINO) focus on modeling intra-speaker variability but are more sensitive to training conditions.

Finally, the performance of SSL frameworks is evaluated on in-domain and out-of-domain benchmarks for SV. The best results are obtained with SimCLR and DINO, achieving 6.41\% and 2.82\% EER on VoxCeleb1-O using the ECAPA-TDNN encoder, while the supervised baseline reaches 1.34\% EER. An extensive comparison of the performance of single- and multi-stage methods from the literature is also provided. The latter demonstrates that numerous methods have been proposed for combining SSL frameworks with different techniques tailored for SV over the last few years. State-of-the-art performance is achieved with methods based on the DINO framework, and multi-stage systems further improve downstream results by generating pseudo-labels to optimize a larger model in a supervised fashion. To highlight the relevance of SSL, a label-efficient evaluation is conducted, showing that using half of the labels is sufficient to match the performance of a supervised model, and that SSL consistently surpasses it in low-resource settings when the amount of labeled data is reduced.

The key limitations of existing work in the field and potential directions for future research are outlined below.
\begin{itemize}
    \item \textbf{Scale.} SSL has shown promising results when applied to vast amounts of unlabeled audio data \cite{baevski2020wav2vec2.0,hsu2021HuBERT,chen2022WavLM}. Therefore, the ability of SSL to learn better speaker representations with larger training sets collected ``in-the-wild'' \cite{yakovlev2023VoxTube,lin2024VoxBlink,lin2024VoxBlink2} and larger encoder architectures remains to be determined.
    \item \textbf{Training set.} The field of SSL for SV has benefited greatly from the existence of VoxCeleb datasets since almost all the existing systems are trained on VoxCeleb2. The application of SSL to other speaker-labeled datasets should be studied in detail, including the effect of using training data created under different conditions and with less channel-related variability.
    \item \textbf{Channel information.} The main bottleneck of SSL is the resulting reliance on channel-related information, which is encoded as much as speaker identity in the representations. This issue arises because of the inability of the data-augmentation module and the positive sampling strategy to create anchor-positive pairs that simulate the diversity of training samples among a given speaker class. Several methods were proposed in the literature to improve data-augmentation \cite{kang2022LMix}, the positive sampling \cite{tao2023DPP,lepage2025SSPS}, or by introducing regularization techniques \cite{huh2020AAT}. However, these solutions should be explored in greater depth, as creating relevant anchor-positive pairs is fundamental for guiding SSL frameworks to learn the most appropriate information for the downstream task. This is considered a key factor in bridging the gap with supervised performance, enabling SSL to achieve comparable levels of accuracy and robustness.
    \item \textbf{Prototypes.} The other main difference between supervised and SSL systems is the ability of the former to map each training sample to a prototype of the corresponding speaker class. This training paradigm enables all samples of a class to be mapped to a single learnable representation, leading to a more effective training objective. Additionally, it allows for the application of constraints, such as margins (e.g., AAM-Softmax loss), to further separate inter-speaker representations in the latent space, which can significantly benefit SV.
    \item \textbf{Projector and hyperparameters.} The effect of enabling and disabling the projector has been investigated in this work, but a more in-depth study would be highly valuable. In particular, the impact of its architecture, the hidden and output dimensions, and the information that is discarded at the representation level. Furthermore, the role of other hyperparameters, such as the temperature and momentum coefficient, remains not well understood and deserves further investigation.
    \item \textbf{Multi-modalities.} Speaker identity attributes are embedded across various modalities, from audio (speech) to image (face) domains. SSL could benefit from multi-modal learning strategies that leverage both speech and visual information to enhance representation quality and robustness. Several works have already explored this topic and demonstrate very promising results when relying on VoxCeleb video data \cite{tao2023DPP,cai2022IncorporatingVisualInformation,han2024CADINO}.
    \item \textbf{Other downstream tasks.} Other information related to the speaker, such as language, emotion, age, and gender, is captured by speaker representations learned with SSL \cite{ravanelli2020PASEPlus,fan2021ExploringWav2vec2.0SV,cho2022DINO}. The research community could investigate the extent to which SSL can be tailored to each specific downstream task by disentangling the information in the representations.
\end{itemize}

SSL has made tremendous progress across various fields, notably in the context of speech representation learning for ASR. However, further research is needed to explore its application to other speech-related downstream tasks beyond the use of a general large-scale SSL speech foundation model. SSL is considered essential for learning more accurate and robust speaker representations, as it enables models to leverage vast amounts of unlabeled data, thereby improving generalization and performance across diverse real-world conditions. This article aims to serve the community by providing detailed insights into how these methods operate and highlighting key considerations for improving their performance in the future.

\section{Acknowledgements}

This work was performed using HPC resources from GENCI-IDRIS (Grant 2023-AD011014623) and has been partially funded by the French National Research Agency (project APATE - ANR-22-CE39-0016-05).

\bibliographystyle{elsarticle-num} 
\bibliography{biblio,biblio_additional}

\end{document}